\documentclass[aps,prc,showpacs,preprint,nofootinbib]{revtex4}

\usepackage{epsfig}
\usepackage{graphicx,amsmath}
\newcommand\ba{\begin{eqnarray}}
\newcommand\ea{\end{eqnarray}}

\newcommand{\be}{\begin{equation}}
\newcommand{\ee}{\end{equation}}

\newcommand{\bas}{\begin{eqnarray*}}
\newcommand{\eas}{\end{eqnarray*}}

\begin{document}
\title{\bf \large Search for Low Mass Exotic Baryons in One Pion
Electroproduction Data Measured at JLAB}
\author{
B. Tatischeff}

\altaffiliation{e-mail : tati@ipno.in2p3.fr}

\affiliation{\it
$^1$Institut de Physique Nucl\'eaire\\CNRS/IN2P3, F--91406 Orsay Cedex, France\\
$^2$Univ Paris-Sud, Orsay, F-91405
 }

\author{E. Tomasi-Gustafsson}
\altaffiliation{e-mail : etomasi@cea.fr}
\affiliation{\it DAPNIA/SPhN, CEA/Saclay, 91191 Gif-sur-Yvette Cedex, 
France}

\date{\today}

\vspace{0.5cm}

\vspace*{1cm}
\begin{abstract}
This paper aims to give further evidence for the existence of low mass exotic baryons. Narrow structures in baryonic missing mass or baryonic invariant mass were previously observed during the last ten years. Since their existence is sometimes questionable, the structure functions of one pion electroproduction cross sections,
measured at JLAB, are studied to add informations on the possible existence of these
narrow exotic baryonic resonances.
\end{abstract}
\maketitle
\section{Introduction}
This paper is dedicated to a reanalysis of existing data on one pion electroproduction cross sections measured at JLAB.
Although observed in several reactions and in different kinematical conditions,
the narrow low mass baryonic structures are sometimes considered with skepticism. Indeed a few dedicated experiments were not able to observe them. This result will be discussed below. In order to disentangle that situation, it is necessary to study new
 data
obtained with a fairly good resolution. The one pion electroproduction cross sections,
 measured at JLAB, are, in principle, appropriate for such study. This study is significant since these
 structures - if any - will be exotic.
 Several reasons plead in favor of their exoticism:\\
- their widths, typically of the order of FWHM$\approx$10 to 20~MeV are much smaller
 than the widths of PDG (Particle Data Group) $N^{*}$ or $\Delta$ resonances,\\
- the first resonances have a mass lower than the pion production threshold mass,\\
- there is no room for these resonances, in the mass range discussed here, within the
 many-quark models for baryons if we consider only $qqq$ configurations \cite{caps}.\\
\hspace*{4.mm}In the second section, the results for narrow low mass baryonic
structures, mainly observed in SPES3 (Saturne), are recalled. Their
 masses are compared to a careful scrutiny of many different data,
obtained by different collaborations, for different physical studies, with
hadronic as well as leptonic probes. The cross sections of the structure functions, 
from backward $\pi^{0}$
 electroproduction on protons, measured by the Hall A Collaboration \cite{lave2}
are considered in sections~III.A. Data on the structure functions
from the ep$\to$e'n$\pi^{+}$ reaction, measured by the CLAS Collaboration \cite{egiy}
in Hall B, 
are discussed in section~III.B. The results of the present analysis are discussed in section IV, and the paper is concluded in section~V.

As mentioned above, narrow structures were observed more often in experiments using incident hadrons than with incident leptons. In the spectra of the 
reaction ep$\to$e'$\pi^{+}$X$^{0}$ studied \cite{jiang} at JLAB (Hall A), no
 significant signal was observed in the range 0.97$\le$M$_{X^{0}}\le$1.06~GeV.
 In a high-resolution experiment studied at MAMI \cite{kohl}, no narrow nucleon resonance below pion threshold was observed in the H(e,e'$\pi^{+}$)X or
D(e,e'p)X reactions. No low-lying exotic baryons (at masses M=1004 and 1044~MeV) were observed at TRIUMF \cite{zoln}  in a double radiative pionic capture on hydrogen.
These three dedicated experiments looked at narrow baryonic structures with
masses below the pion production mass. The absence of signal in these experiments using incident leptons, can be related to
the fact that these narrow structures may have a small coupling to nucleon, their excitation being favored for reactions involving two baryons (for example excitation through intermediate dibaryons).

The lack of observation of narrow baryonic structures below pion production mass, constitutes a further reason to look at data obtained with incident leptons, concerning the mass range above pion production threshold.
\section{Recall of already published data showing the existence of narrow low mass baryons}
Previous experiments, performed at SPES3 (Saturne), thanks to good resolution and high
 statistics, exhibit narrow structures in different 
hadronic masses. Only results concerning baryons will be discussed here.  Two reactions: 
\be
p+p\to p+p+X 
\label{eq:eq1}
\ee
and
\be
p+p\to p+ \pi^++X 
\label{eq:eq2}
\ee
were studied.  Structures were observed in the missing mass M$_{X}$ of reaction (\ref{eq:eq2}) \cite{bor1} and in the invariant mass M$_{pX}$ of reaction (\ref{eq:eq1}) and  in the invariant masses M$_{p\pi^{+}}$ and M$_{\pi^{+}X}$ of reaction (\ref{eq:eq2}) \cite{bor2}. The observation in different conditions (reaction, incident energy, spectrometer angle, or observable) at the same mass (within $\pm$3~MeV) was considered a confirmation of their existence. This is summarized in Fig.~1 of Ref. \cite{bor2} in the mass range 
1.0$\le$M$\le$1.4~GeV. In the figure, columns (a) to (f) 
correspond to different variables or incident energies of reaction (2), columns (g) and (h)
 correspond to reaction dp$\to$ppX at two different incident energies \cite{bor2}, column (i) 
describes data from $\gamma$n$\to$p$\pi^{-}\pi^{0}$ reaction studied at MAMI \cite{zab}
 and column (j) to data from $\gamma$p$\to\pi^{+}$n reaction studied at Bonn \cite{dann}.
The narrow structures masses observed are: 1004, 1044, 1094, 1136, 1173, 1249, 1273,
 1339, and 1384~MeV.
 
Additional signatures of narrow baryonic structures, were observed either in dedicated experiments or  extracted from cross sections obtained and published by different authors studying other problems. They are quoted in \cite{bor1} and  \cite{bor2} and will not be recalled here.
 
Precise spectra of the  p($\alpha$,$\alpha$')X reaction were obtained at SPES4
(Saturne) in order to study the radial excitation of the nucleon in the
P$_{11}$(1440~MeV) Roper resonance. The measurements were done using a
T$_{\alpha}$=4.2~GeV incident beam. The spectrum at $\theta_{\alpha'}$=0.8$^{0}$
was published in \cite{mor1} and the spectrum at $\theta_{\alpha'}$=2$^{0}$ was published
in \cite{mor2}. A first large peak around M$_{X}\approx$1130~MeV
($\omega\approx$240~MeV), was associated
to the projectile excitation, and a second large peak around
M$_{X}\approx$1345~MeV ($\omega\approx$510~MeV), was associated
to the target excitation. Above them lie narrow peaks \cite{bor3}, characterized
by a large number of standard deviations, since the highest channel at
$\theta$=0.8$^{0}$ contains approximately 2.5$\times$10$^{4}$ events 
 (see Fig.~2 and 3  ). These structures were not discussed by the authors. A detailed 
discussion of the spectrometer and of the detection performances was given in \cite{bor3}
and will not be repeated here as well as the checks performed and the final
precision obtained. Figs.~2 and 3 show the spectra. Table 1 gives the correspondance among the letters naming the structures, their masses and the masses of the corresponding structures
extracted from SPES3 data. Such correspondance is illustrated in Fig.~4.

Finally we list, in the mass range studied here: 
1.1 $\le$ M $ \le$ 1.56~GeV, eight narrow  masses at M=1136, 1173, 1249, 1273, 1339, 1384, 1480, and 1540~MeV. A well separated 
structure at M=1479~MeV was extracted from pp$\to$ppX and pp$\to$pp$\pi^{0}$
reactions \cite{bor2}. Several other structures were extracted in the same work, 
at larger masses, but with a too small separation to be seen in the two
 lower resolution experiments of one pion electroproduction on proton, considered in the present work. 
\section{Analysis of one pion electroproduction structure function cross sections measured at JLAB}

The reactions $\gamma^*+p\to \pi^0+p$  and $\gamma^*+p\to \pi^++n$ have been measured at different kinematical conditions.

The differential cross sections are expressed by the following equation \cite{guich}:\\
\be
d^{2}\sigma/d\Omega_{p} =  d^{2}\sigma_{T}/d\Omega_{p}  +  \varepsilon d^{2}\sigma_{L}/d\Omega_{p} + \surd[2\varepsilon(1+\epsilon)] 
d^{2}\sigma_{LT}/d\Omega_{p}   
\cos(\Phi) + \varepsilon d^{2}\sigma_{TT}/d\Omega_{p} \cos(2\Phi).
\label{eq:eqs}
\ee

The $\sigma_{T}$, $\sigma_{L}$, $\sigma_{TL}$, and $\sigma_{TT}$ structure
 functions are bilinear combinations of the helicity amplitudes, depending
 only on the variables Q$^{2}$, W, and $\theta$. $d^{2}\sigma_{T}$ is the transverse part of
 the cross section, $d^{2}\sigma_{L}$ is the longitudinal part of the cross section, and
 $d^{2}\sigma_{TL}$ and $d^{2}\sigma_{TT}$ are interference parts.\\
\hspace*{4.mm} 
$\theta$ is the polar angle between initial and final protons, in the CM system defined
 by the final proton and missing particle $(\bar{q}+\bar{p})$.
 $\varepsilon$=[1+2(1+$\nu^{2}/Q^{2}$)tan$^{2}\theta_{e}/2]^{-1}$ is the polarization
parameter or the virtual photon polarization. $\nu$=E$_{i}$ - E$_{f}$ is the energy transfer.
Q$^{2}$=4~E$_{i}$E$_{f}$sin$^{2}\theta_{e}$/2=$(\bar{k}_{e}-\bar{k'}_{e})^{2}$ is the
 four momentum transfer squared. M is the proton mass and 
W=[M$^{2}$ + 2M$\nu$ - Q$^{2}]^{1/2}$ is the mass of the hadronic system. The structure functions
are plotted versus W. $\Phi$ is the azimuthal angle between the leptonic and the
 hadronic planes.
 
The structure functions $\sigma_{TT}$, $\sigma_{TL}$, and the linear combination $\sigma_{T} + \epsilon \sigma_{L}$ were obtained by fitting the $\Phi$ dependence of
 the cross section to a function of the form:\\
\ba
 F(\theta)=A+B\cos(\Phi)+C\cos(2\Phi).
\label{eq:eqff}
\ea
They are described
theoretically by the phenomenological MAID model \cite{maid}.
 The MAID model uses an effective Lagrangian approach to calculate the Born
background, including $\omega$ and $\rho$ meson calculations. The
 background is unitarized in the K-matrix approximation. The resonant
 amplitudes are determined by fitting the world pion production data.
The MAID2003 model is a fit to predominantly $\pi^{0}$p channel.\\
\hspace*{4.mm}The MAID calculations describe the main shapes of the structure functions, especially at small $\theta$ angles. 

We compute  MAID using all baryonic resonances from $P_{33}$(1232) up to $F_{37}$(1950)
 \cite{maid}, although the last ones lie outside the range of study. The calculation holds
 background, resonances and interferences between both.
 
MAID contains the experimental phase shifts and includes most of the known physics, in electroproduction processes. Therefore, in order to enhance the physics which is not contained in MAID, we build the difference between the experimental spectra and the MAID results. The figures show that such difference exhibits narrow structures located at similar masses. Then, we fit the results of the subtraction of data from MAID. Interferences 
between these small eventual structures, and the background and broad resonances exist.
Since the amplitudes of broad PDG resonances (and Born background) vary little in the smaller range of each narrow resonance, the extracted position of these narrow resonances should be not much affected by these
interference terms. Such effect exist in all experiments since these resonances lie always above other physics and background. However it was observed, see Fig.~1, that the masses observed were quite stable.\\
\hspace*{4.mm}
The fits shown are obtained using gaussians and masses given in previous section, namely
 those extracted from previous experiments. No attempt is done to adjust any mass. The
 width also is taken,
 arbitrarily, to be $\sigma$=24~MeV, in both reactions, without any attempt to adjust it.   
The experimental statistical errors are kept as errors for the data minus MAID values. In some cases, the large error bars prevent to give firm confidence on the peak extractions which 
were only done in view of consistency. 

In these experiments, the masses below pion production cannot be observed. Also the 
structure at M=1094~MeV is missing since the data start at M=1110~MeV.\\
\hspace*{4.mm}When $\sigma_{TT}$ and $\sigma_{TL}$ structure functions are 
interference terms, with possible positive and negative values, the third structure function:
$\sigma_{T}~+~\epsilon~\sigma_{L}$ is positive, since obtained by the square of 
amplitudes.  This is the case for data and for MAID results. However the difference between both
 involve interference terms between "classical amplitudes", described by MAID, and new
 amplitudes from narrow baryons which existence is discussed. Therefore, after the 
difference, the result of  this structure function $\sigma_{T}~+~\epsilon~\sigma_{L}$
can be negative. We suppose that the amplitudes of the Born background and broad
resonances vary slowly in the range of narrow resonances. This justifies, as already pointed out, the statement that 
our procedure should exhibit narrow peaks, if any, close to their genuine masses. 
\subsection{The $\gamma^{*}$p$\to~\pi^{0}$p reaction}
The backward cross sections of the structure functions of the $\gamma^{*}$p$\to~\pi^{0}$p
 reaction
were measured by the Hall A Collaboration \cite{lave2} at four angles $\theta$ and at
 Q$^{2}$=1~GeV$^{2}$.
\subsubsection{The $\sigma_{T}~+~\epsilon\sigma_{L}$ structure function}
Fig.~5 shows the $\sigma_{T}~+~\epsilon\sigma_{L}$ structure function at $\theta$=167.16$^{0}$
 in inserts (a) and (b) and at $\theta$=157.67$^{0}$ in inserts (c) and (d). Inserts (a) and (c) show
 the data (full circles) the MAID results (dashed curves) and the difference (full stars). Inserts (b) and
(d) show the previous difference (full circles), the peaks for individual masses and the total
 spectrum obtained with assumption of no interference. At both angles a large peak is observed at 
a somewhat larger mass (M$\approx$1200~MeV instead M=1173~MeV).  A narrow structure at M$\approx$1540~MeV is well defined in insert (b). Fig.~6 shows the corresponding results for $\theta$=155.05$^{0}$ in inserts (a) and (b) and at $\theta$=145.59$^{0}$  in inserts (c) and (d).  We observe that the quality of the fit with MAID, at backward angles, gets spoiled when the angles decrease.
 
Fig.~7 shows the angular variation versus $\theta$ of the yield of the seven narrow masses
as extracted from Figs.~5 and 6. The error bars here are arbitrarily put to 20$\%$ of the height of each peak, increased by 10$\%$ of the highest peak in order to get a reasonable error for  small peaks. We observe
a smooth and continuous variation  of the observed yields. The curves are a tentative fit of the points by the function 
\be
F=\sum_{i=0}^{2}a_{i}\cos^{2i}\theta.
\label{eq:eqf}
\ee 
In the small angular range covered by the experiment, all seven  peaks show a similar behavior. We do not attempt to draw conclusions on the spin and parity of the narrow resonances from the present, restricted angular distributions. 
\subsubsection{The $\sigma_{TT}$ structure function}
Fig.~8 shows the results for the $\sigma_{TT}$ structure function at $\theta$=167.16$^{0}$ in inserts (a)
 and (b), and at $\theta$=157.67$^{0}$ in inserts (c) and (d). Here also the data, the 
MAID results, the difference
 between both, and the fits of the peaks are defined as previously for the  
$\sigma_{T}~+~\epsilon\sigma_{L}$ structure function. Fig.~9 shows the results for the $\sigma_{TT}$ 
structure function at $\theta$=151.05$^{0}$ in inserts (a) and (b), and at $\theta$=145.59$^{0}$ in
 inserts (c) and (d). Fig.~10 shows the angular variation of the seven $\sigma_{TT}$  structure functions
corresponding to the seven peaks studied. In all inserts the yield for the largest angle is close to zero.
A rather good continuity is observed, except in insert (a) which corresponds to M=1136~MeV narrow structure peak. The same function as previously, Eq (\ref{eq:eqf}),  is used for the fits.
\subsubsection{The $\sigma_{TL}$ structure function}
Fig.~11 shows the results for the $\sigma_{TL}$ structure function at $\theta$=167.16$^{0}$ in inserts (a)
 and (b), and at $\theta$=157.67$^{0}$ in inserts (c) and (d). Fig.~12 shows the results for the $\sigma_{TL}$ 
structure function at $\theta$=151.05$^{0}$ in inserts (a) and (b), and at $\theta$=145.59$^{0}$ in
 inserts (c) and (d). In this structure function, the fit with MAID is poor at all angles; at the smallest angle $\theta$=145.59$^{0}$, the peak at M=1210~MeV reaches only 
25$\%$ of its experimental value.
Fig.~13 shows the angular variation of the seven $\sigma_{TL}$ structure functions, corresponding to the seven peaks studied.  We observe the 
continuous behavior of the other distributions, again fitted with the same function versus
 $\cos \theta$. Here again, the small angular range prevents to give strong
 importance to these fits which are merely an indication of similar shapes between all peaks.
\subsection{The $\gamma^{*}$p$\to~\pi^{+}$n reaction}
The $ep\to e'n \pi^{+}$ reaction was measured at JLAB in Hall B, by the CLAS Collaboration \cite{egiy}. These data are less precise than those discussed above from Hall A. The cross sections of the three structure functions were extracted at four values of four momentum transfers: Q$^{2}$=0.3, 0.4, 0.5, and 0.6 GeV$^{2}$, and at ten angles $\theta$=7.5$^{0}$,
22.5$^{0}$, 37.5$^{0}$, 52.5$^{0}$, 67.5$^{0}$, 82.5$^{0}$, 97.5$^{0}$, 112.5$^{0}$, 127.5$^{0}$, and
142.5$^{0}$. We analyze here the difference between data and MAID calculations for two Q$^{2}$ values
Q$^{2}$=0.3 and 0.4~GeV$^{2}$, and six center of mass angles $\theta$=7.5$^{0}$, 
22.5$^{0}$, 52.5$^{0}$, 67.5$^{0}$, 97.5$^{0}$, and 127.5$^{0}$. The data at  $\theta$=142.5$^{0}$
are very imprecise. No data are available at Q$^{2}$=0.6~GeV$^{2}$ for the hadronic system mass W larger
than W=1.4~GeV. As previously discussed, the figures illustrate the experimental structure functions, the MAID results and their differences on the left side, whereas, on the right side these differences are plotted with the corresponding fits.
\subsubsection{The $\sigma_{T}~+~\epsilon\sigma_{L}$ structure functions} 
Fig.~14 exhibits the  $\sigma_{T}~+~\epsilon\sigma_{L}$ structure functions at $\theta$=7.5$^{0}$ and
Q$^{2}$=0.3~GeV$^{2}$ in inserts (a) and (b), and at $\theta$=7.5$^{0}$ and
Q$^{2}$=0.4~GeV$^{2}$ in inserts (c) and (d). For this structure function, and at such small angle, MAID describes fairly well the data up to W=1350~MeV, however a difference remains which can be fitted reasonably well 
with the same masses as before. The next figures Fig.~15, Fig.~16, Fig.~17, Fig.~18, and Fig.~19, 
correspond respectively to the same analysis but different angles $\theta$=22.5$^{0}$, 52.5$^{0}$, 
67.5$^{0}$, 97.5$^{0}$, and 127.5$^{0}$. Some masses, i.e. at M=1173~MeV, 1210~MeV and 
1480~MeV are better defined than the others. Here again the cross sections at Q$^{2}$=0.3~GeV$^{2}$ 
and  Q$^{2}$=0.4~GeV$^{2}$ are not very different. Fig.~20 shows the angular variations of the seven
 $\sigma_{T}~+~\epsilon\sigma_{L}$ structure functions, corresponding to the seven narrow structure
 masses studied. The error bars are again defined as explained above for the first reaction. Full circles correspond to Q$^{2}$=0.3~GeV$^{2}$ and empty circles correspond to
Q$^{2}$=0.4~GeV$^{2}$. The data corresponding to both four-momentum transfers are close, most of the time.
Full curves correspond to tentative fits with the function defined above, Eq. \ref{eq:eqf}; in one case, insert (a), the fit is obtained with an odd function of $\theta$, namely $\sin 2\theta$.
\subsubsection{The $\sigma_{TT}$ structure functions} 
Fig.~21 shows the results for $\sigma_{TT}$ structure function at $\theta$=7.5$^{0}$, in inserts (a) and (b) for Q$^{2}$=0.3~GeV$^{2}$ and in inserts (c) and (d) for Q$^{2}$=0.4~GeV$^{2}$ data.
At this angle, there is no precise experimental data in the mass range 1230$\le$M$\le$1340~MeV.                            
The description by MAID fails totally to describe the data. The results for $\theta$=22.5$^{0}$ are shown in Fig.~22, inserts (c) and (d) for Q$^{2}$=0.3~GeV$^{2}$ and in Fig.~23, inserts (c) and (d)
for Q$^{2}$=0.4~GeV$^{2}$ data. Fig.~24 show the results of the $\sigma_{TT}$ structure function
at Q$^{2}$=0.3~GeV$^{2}$. Inserts (a) and (b) show the results obtained at $\theta$=52.5$^{0}$, and inserts (c) and (d) show the results at $\theta$=67.5$^{0}$. Fig.~25 shows the results identical to those from Fig.~24, but for Q$^{2}$=0.4~GeV$^{2}$. Fig.~26 shows the results for 
$\sigma_{TT}$ structure function at $\theta$=97.5$^{0}$ at Q$^{2}$=0.3~GeV$^{2}$ in inserts (a) and (b) and at Q$^{2}$=0.4~GeV$^{2}$ in inserts (c) and (d). The error bars decrease with increasing angles, and consequently the structure's definitions are good. Fig.~27 shows the 
$\sigma_{TT}$ structure function at $\theta$=127.5$^{0}$ at Q$^{2}$=0.3~GeV$^{2}$ in inserts (a) and (b) and at Q$^{2}$=0.4~GeV$^{2}$ in inserts (c) and (d).\\
\hspace*{4.mm}Fig.~28 shows the angular variation of the $\sigma_{TT}$ structure function with an attempt to fit the data with a low order polynomial of $\cos^{2n}\theta$.
The full circles correspond to Q$^{2}$=0.3~GeV$^{2}$, and the empty circles correspond to Q$^{2}$=0.4~GeV$^{2}$.
We observe that in several inserts, corresponding to different narrow mass structures, 
the fitted curves describe very satisfactorily most of the data.
\subsubsection{The $\sigma_{TL}$ structure functions} 
Fig.~29 shows the cross-section of the $\sigma_{TL}$ structure function at $\theta$=7.5$^{0}$.
Inserts (a) and (b) correspond to Q$^{2}$=0.3~GeV$^{2}$, inserts (c) and (d) correspond to Q$^{2}$=0.4~GeV$^{2}$. Here also, as it was the case at the same angle for the 
$\sigma_{TT}$ structure function, there is no precise experimental data in the mass range around
M$\approx$1300~MeV. The results at 22.5$^{0}$ are shown in Fig.~22 inserts (a) and (b) for 
Q$^{2}$=0.3~GeV$^{2}$, and in Fig.~23 inserts (a) and (b) for Q$^{2}$=0.4~GeV$^{2}$.
Fig.~30 shows the results for Q$^{2}$=0.3~GeV$^{2}$ $\theta$=52.5$^{0}$ in inserts (a) and (b),
and $\theta$=67.5$^{0}$ in inserts (c) and (d). The peak's extraction from insert (b) is meaningless,
it is only shown for sake of consistency. Fig.~31 shows the results corresponding to the same
angles but for  Q$^{2}$=0.4~GeV$^{2}$. Here again the $\theta$=52.5$^{0}$ data do not allow to extract peaks. Fig.~32 shows the results for $\theta$=97.5$^{0}$  Q$^{2}$=0.3~GeV$^{2}$ in inserts (a) and (b), and Q$^{2}$=0.4~GeV$^{2}$ in inserts (c) and (d). Fig.~33 shows the 
results for $\theta$=127.5$^{0}$  Q$^{2}$=0.3~GeV$^{2}$ in inserts (a) and (b), and Q$^{2}$=0.4~GeV$^{2}$ in inserts (c) and (d).\\
\hspace*{4.mm}Fig.~34 shows the angular variations of the  $\sigma_{TL}$ structure function. The insert (a) \hspace*{4.mm}(M=1136~MeV) is fitted with an odd function,  $1.6\sin 2\theta$; the insert (b) (M=1210~MeV) is fitted with the function $4.5\cos \theta$. The other curves are fits to the points with a low order polynomial of 
$\cos^{2n}\theta$. 
\section{Discussion}
Whereas clear peaks are observed in both sides of the mass range studied here, 
this is not always the case in the mass range M$\approx$1.4~GeV. No attempt to get a better adjustment by shifting the masses, is done on the fits shown above. Most of the extracted structure function surfaces, exhibit a smooth  angular 
variation. This result justifies, {\it a posteriori}, the attempt to associate the difference
 between structure functions and MAID, with the existence of narrow 
baryonic structures.

One argument not to attribute the differences described above to 
deficiencies in MAID, lies in the smallness of the widths of the residual peaks.
Indeed, we expect that an eventually poor description of the data by MAID would result in broader effects.
\subsection{Possible isospin values for the narrow structures}
Both reactions were studied at complementary angles, roughly
in the range 0$\le\theta\le$140$^{0}$ for the $\gamma^{*}p\to\pi^{+}$n reaction and
 140$\le\theta\le$170$^{0}$ for the $\gamma^{*}p\to\pi^{0}$p reaction. However in 
both reaction, the four momentum transfer is different. We observe small variations 
of  the cross section  between Q$^{2}$=0.3~GeV$^{2}$ and 
Q$^{2}$=0.4~GeV$^{2}$, but an extrapolation up to Q$^{2}$=1~GeV$^{2}$ may not be  justified. Both reactions are related by isospin Clebsh-Gordan coefficients.
If we neglect the  variation of the structure function due to different Q$^{2}$ values, the
intermediate resonance $N^{*}$ with isospin 1/2, will favor  $\pi^{+}n$ by a factor of two,
and the intermediate resonance $\Delta$ with isospin 3/2 will favor  $\pi^{0}$p by
 a factor of two. Figs.~20, 28, and 34 show
the angular variations of the structure functions, respectively 
$\sigma_{T} + \epsilon \sigma_{L}$, $\sigma_{TT}$, and $\sigma_{TL}$, with the 
results of both reactions. 
The results from the $ep\to ep'\pi^{0}$ reaction at backward angles are 
drawn with stars. Both reactions display cross sections with smooth behavior for several inserts. A continuous curve reproducing the behavior  for  both reactions, may be considered as an indication of the excitation of isospin 3/2 resonance. In this case, the increase by a factor of two, due to isospin, could be compensated by a reduction by a similar 
factor  due to the increase in $Q^{2}$ from 0.3-0.4~GeV$^{2}$ up to
1~GeV$^{2}$.

In order to tentatively suggest isospins for the narrow structures, we
apply the following rule: when the structure function of the
 $\gamma^{*}p\to \pi^{0} p$ reaction and the structure function of the 
 $\gamma^{*} p\to\pi^{+} n$ reaction follow the same line, we propose an isospin value of 3/2 for the baryonic structure; when the structure function of the first reaction
is much smaller than the one of the second reaction, we suggest an isospin 1/2.
Table ~2 shows the possible isospin attributions. We observe that the structures at
W=1136~MeV (insert (a)), should have isospin T=1/2, since all three structure functions predict such value. Isospin T=3/2 is predicted twice for M=1210~MeV, M=1277~MeV, and M=1480~MeV. Isospin T=1/2 is predicted twice for M=1339~MeV, M=1384~MeV, and M=1540~MeV. Therefore narrow structures at M=1210~MeV and 
M=1277~MeV could be "substructures" of the broad PDG $\Delta (1232)$ $P_{33}$ resonance; just as the narrow structure at M=1480~MeV could be a "substructure" of the broad PDG $\Delta (1600)$ $P_{33}$ resonance which total width is estimated \cite{pdg} to be as large as 350~MeV. The narrow structures at M=1339~MeV, M=1384~MeV, and M=1540~MeV could be parts of the $N^{*}(1440)$ $P_{11}$ which total width is also estimated \cite{pdg} to be as large as 350~MeV, and (or) part of the $N^{*}(1520)$ $D _{13}$ broad PDG baryonic resonance. 
\subsection{Possible spin values for the narrow structures}
The curves which fit the angular distributions drawn in Figs.~20, 28, and 34 are obtained, besides a few exceptions, 
using low order polynomials of $x=\cos^{2}\theta$. Due to the relative imprecision of the data, and their rather reduced number, the fits presented are not conclusive. It is not possible to identify the angular distributions with theoretical angular distributions \cite{peierls} \cite{gasio} which are given for cross-sections and not for structure functions. Moreover the theoretical angular distributions may be more complicated for increasing J
values of the narrow baryonic resonances spins. Indeed these angular distributions are polynomials of 
$\cos^{2}\theta$, of the order 2J-1. 

The angular distributions of the $\sigma_{T} + \epsilon \sigma_{L}$ structure
function differ from one narrow structure mass to another. Since the data of the M=1277~MeV structure (insert (c))  scatter too much, the fitted curve  may be meaningless. The experimental distributions of the other inserts are
 continuous. 
 
The angular distributions of the $\sigma_{TT}$  structure function  show a smooth behavior for the inserts (a), (b), (c), (e), and (g). 
 
We observe that the distributions of the $\sigma_{TL}$ 
structure functions have the same shape for the masses corresponding to inserts (c), (d), (e), and (f), namely that they all are proportional to $f=x-x^{2}$ (where $x$ stands for $\cos^{2}\theta$) with different translations. The data are continuous for the inserts (a), (b), (d), (e), and (g). 

Concluding this discussion, we observe that no spin attribution can be made, and only the comparison between $\pi^{0}$ and $\pi^{+}$ electroproduction, may  eventually allow to suggest isospin values.
\section{Conclusion}
The paper presents a contribution to the study of narrow exotic low mass baryons. Above pion production threshold, several narrow baryonic structures could be extracted from experiments using incident leptons. Several results of this kind were presented in \cite{bor2} and were not  recalled here.
The two discussed experiments, namely one pion electroproduction on proton, 
were performed with another aim; their results were then not obtained with 
an appropriate resolution. 

We have shown that the description of the measurements
with MAID was sometimes more qualitative than quantitative. The difference between 
data and results from MAID calculations, exhibits narrow peaks, better defined in both sides of the studied range. We have shown that the entire range can be described by 
structures at the masses extracted from previous experiments performed with 
hadrons and previous analysis. We conclude that these data, although they do not contain by themselves an unambiguous signature, they nevertheless increase the confidence in the genuine existence of these narrow baryonic structures. The comparison between both electroproduction reactions, allowed us to suggest possible isospins for these narrow baryonic structures.

Thanks are due to H. Fonvieille for her interest and critical remarks.

\begin{center}
\begin{figure}[!ht]
\scalebox{0.9}[0.9]{
\includegraphics[bb=1 1  530 530,clip,scale=1]{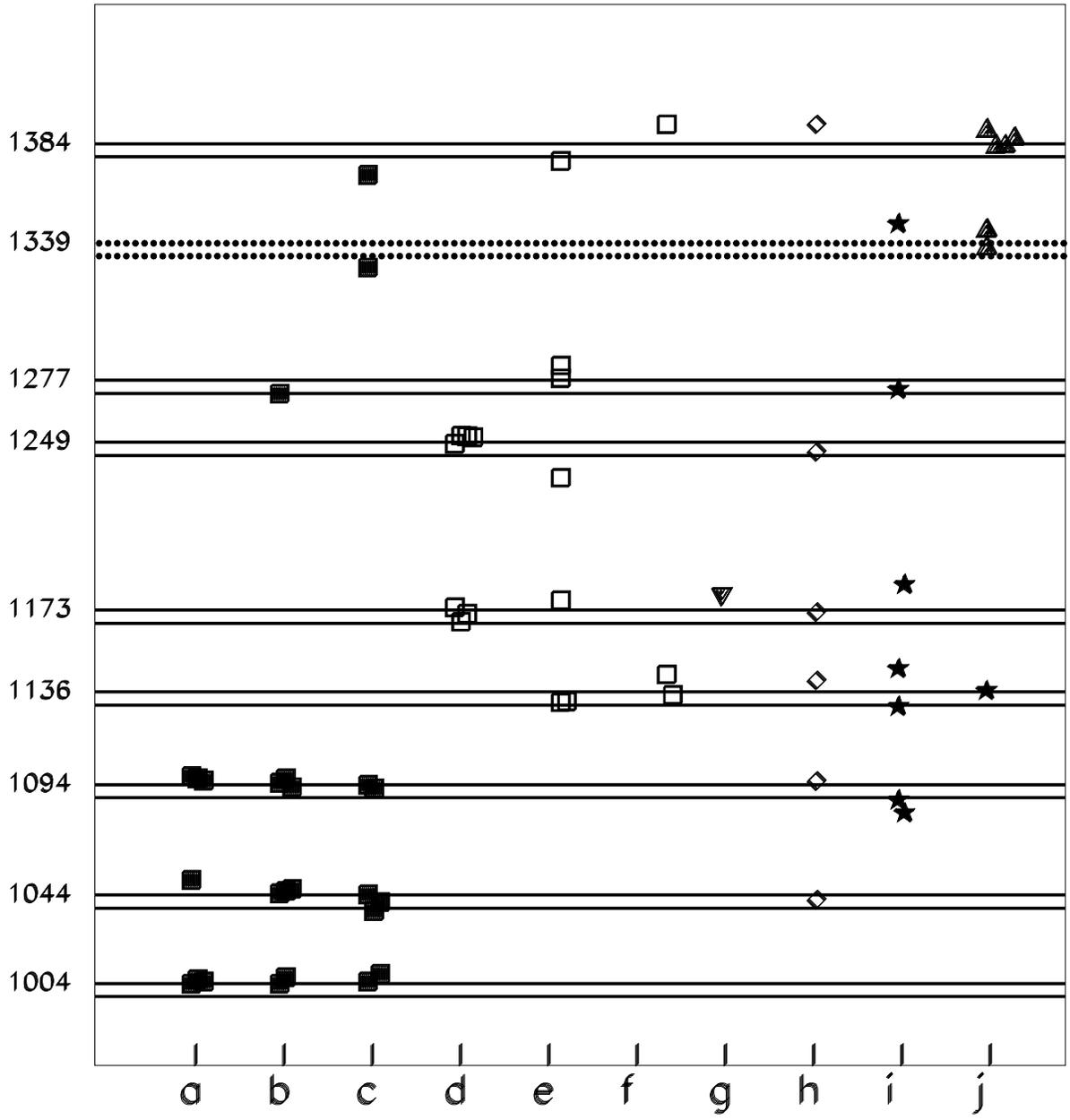}}
\caption [Fig.~1] {Narrow-structure baryonic masses observed in cross-sections from different reactions \protect\cite{bor2}.}
\end{figure}
\end{center}

\begin{figure}[!ht]
\begin{center}
\scalebox{.9}[.9]{
\includegraphics[bb=20 20 530 530,clip=]{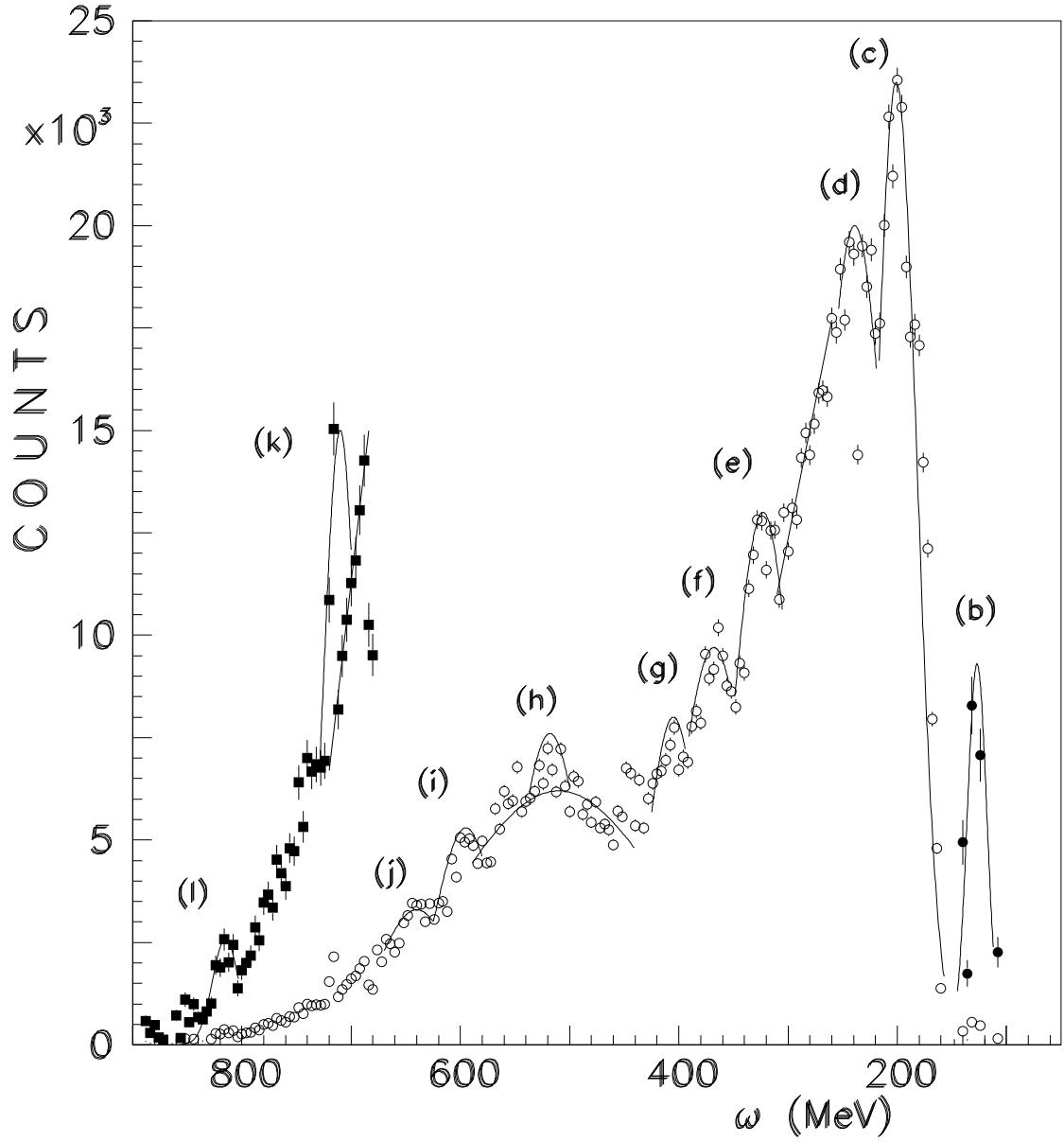}}
\caption [Fig.~2] {Spectra of the p($\alpha,\alpha'$)X reaction studied at SPES4
(Saturne) with T$_{\alpha}$=4.2~GeV and $\theta$=0.8$^{0}$ \protect\cite{mor1}.}
\end{center}
\end{figure}

\newpage\clearpage
\begin{figure}[!ht]
\begin{center}
\scalebox{.9}[.9]{
\includegraphics[bb=20 20 530 530,clip=]{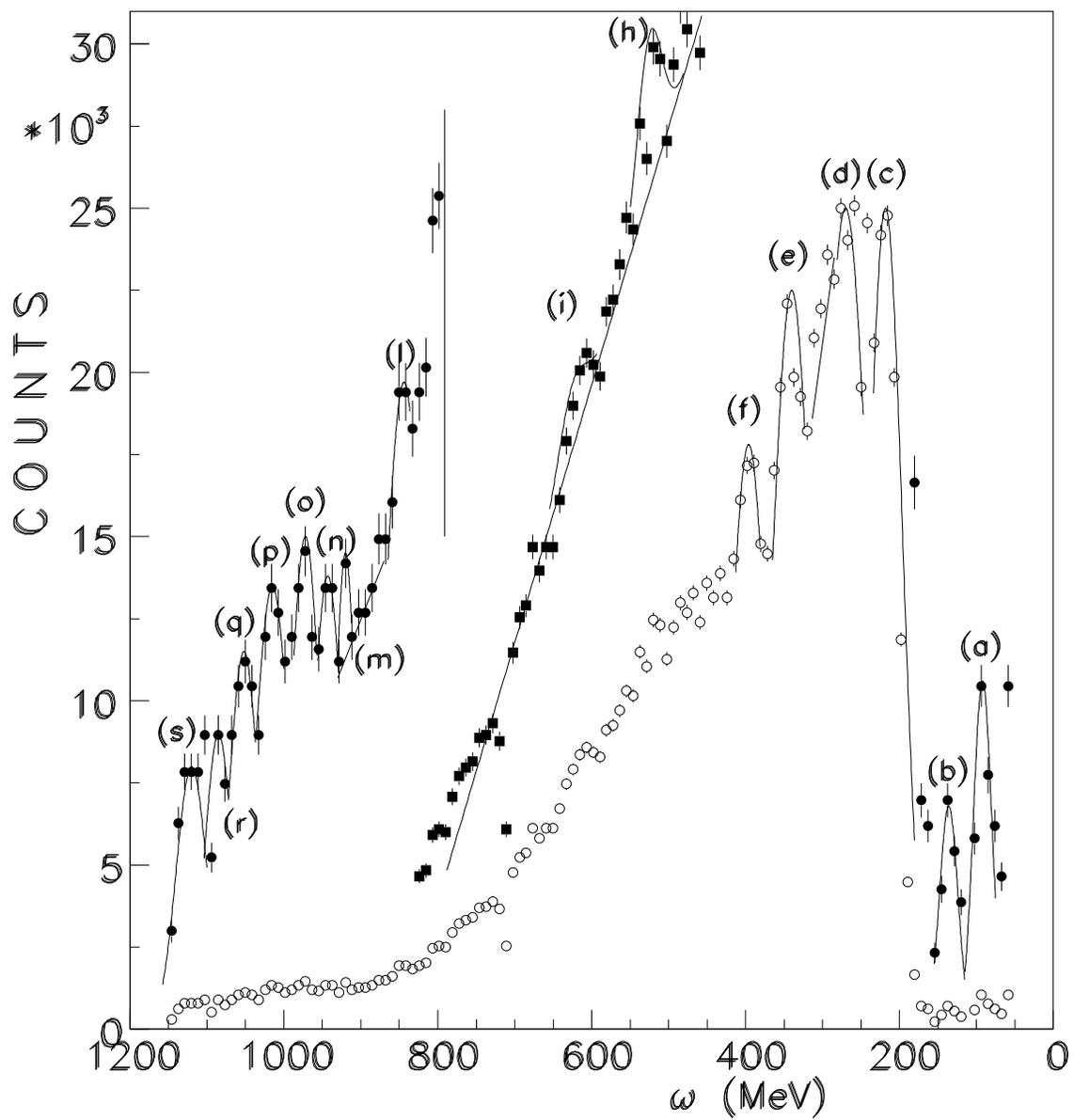}}
\caption [Fig.~3] {Same  as Fig.~2, but obtained at $\theta$=2$^{0}$ \protect\cite{mor2}.}
\end{center}
\end{figure}

\newpage\clearpage
\begin{figure}[!ht]
\begin{center}
\scalebox{.9}[.9]{
\includegraphics[bb=20 20 530 530,clip=]{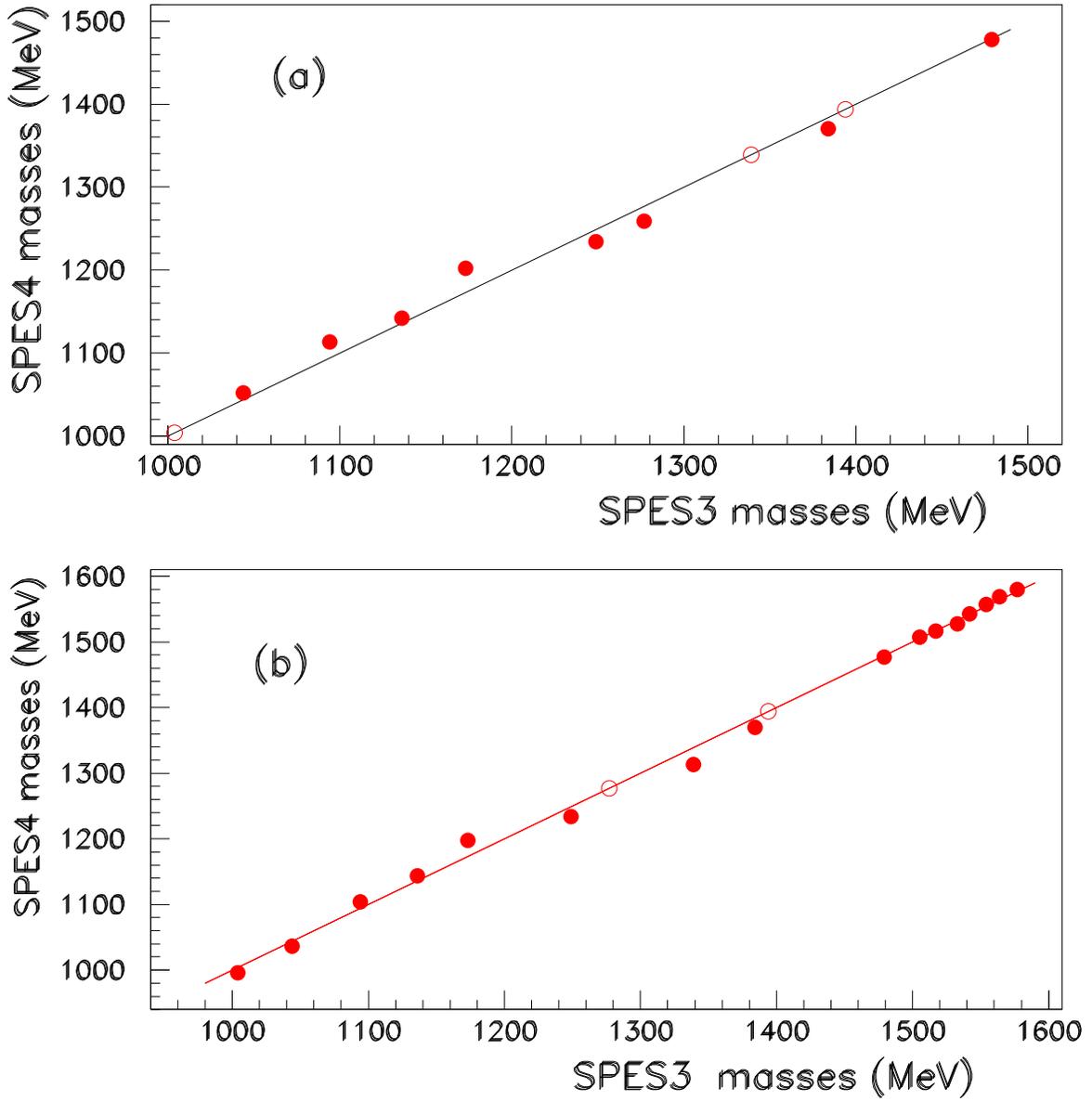}}
\caption [Fig.~4] {Comparison between masses of narrow baryons extracted from
SPES3 and SPES4 data. Inserts (a) and (b) correspond respectively to
$\theta$=0.8$^{0}$ and $\theta$=2$^{0}$. Full circles correspond to narrow structure masses observed in both experiments, empty circles correspond to narrow structure masses observed in only one experiment.}
\end{center}
\end{figure}

\newpage\clearpage
\begin{center}
\begin{figure}[!ht]
\scalebox{0.9}[0.9]{
\includegraphics[bb=12 12  530 550,clip,scale=1]{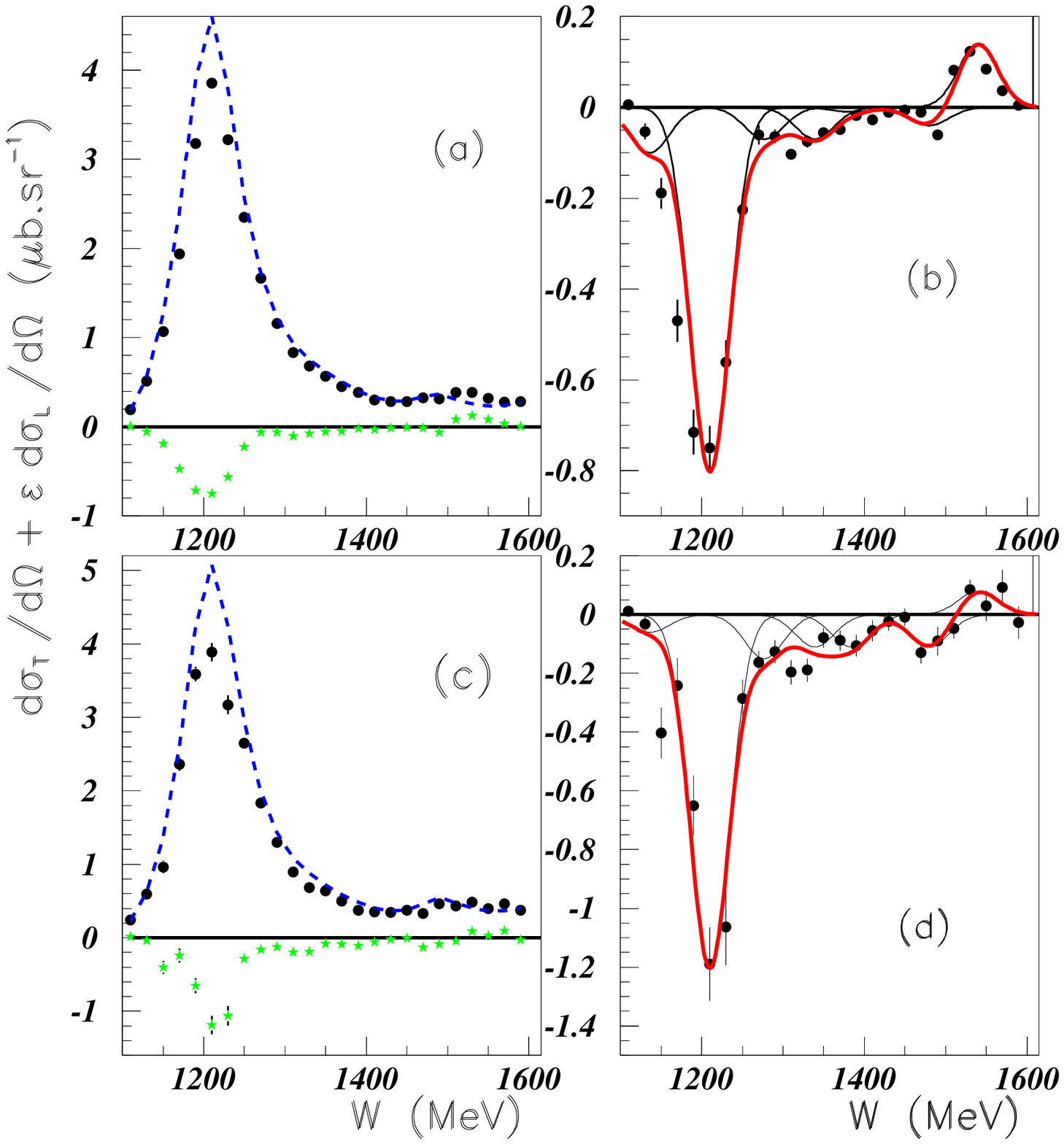}}
\caption [Fig.~5] {$\sigma_{T}~+~\epsilon\sigma_{L}$ structure function of the
$\gamma^{*}$p$\to\pi^{0}$p reaction \protect\cite{lave2}, at $\theta$=167.16$^{0}$
 in inserts (a) and (b) and at $\theta$=157.67$^{0}$ in inserts (c) and (d) (see text). }
\end{figure}
\end{center}

\begin{center}
\begin{figure}[!ht]
\scalebox{0.9}[0.9]{
\includegraphics[bb=12 12  530 550,clip,scale=1]{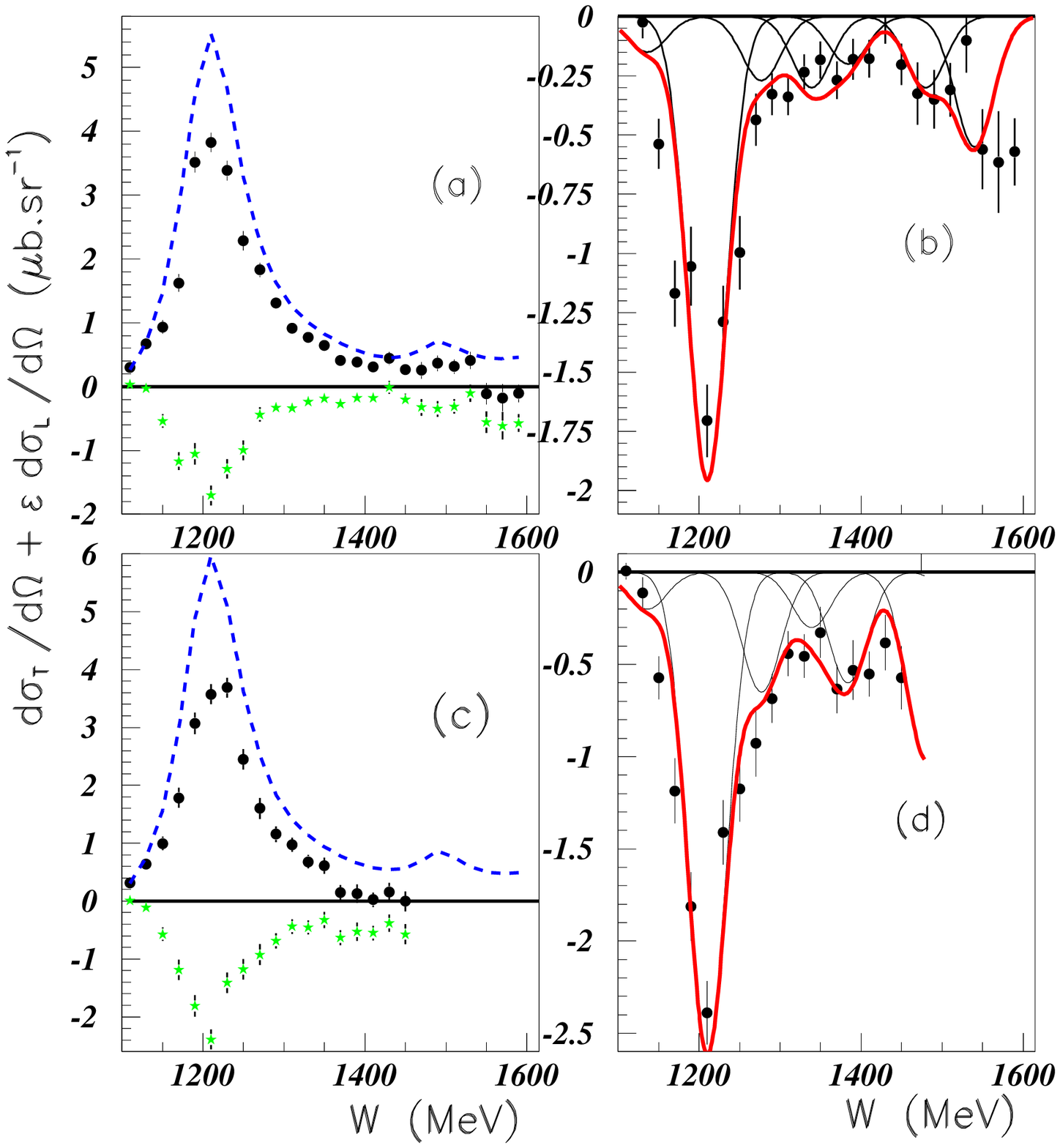}}
\caption [Fig.~6] {$\sigma_{T}~+~\epsilon\sigma_{L}$ structure function of the
$\gamma^{*}$p$\to\pi^{0}$p reaction \protect\cite{lave2}, at $\theta$=151.05$^{0}$
 in inserts (a) and (b) and at $\theta$=145.59$^{0}$ in inserts (c) and (d) (see text). }
\end{figure}
\end{center}

\begin{center}
\begin{figure}[!ht]
\scalebox{0.9}[0.9]{
\includegraphics[bb=12 12  530 450,clip,scale=1]{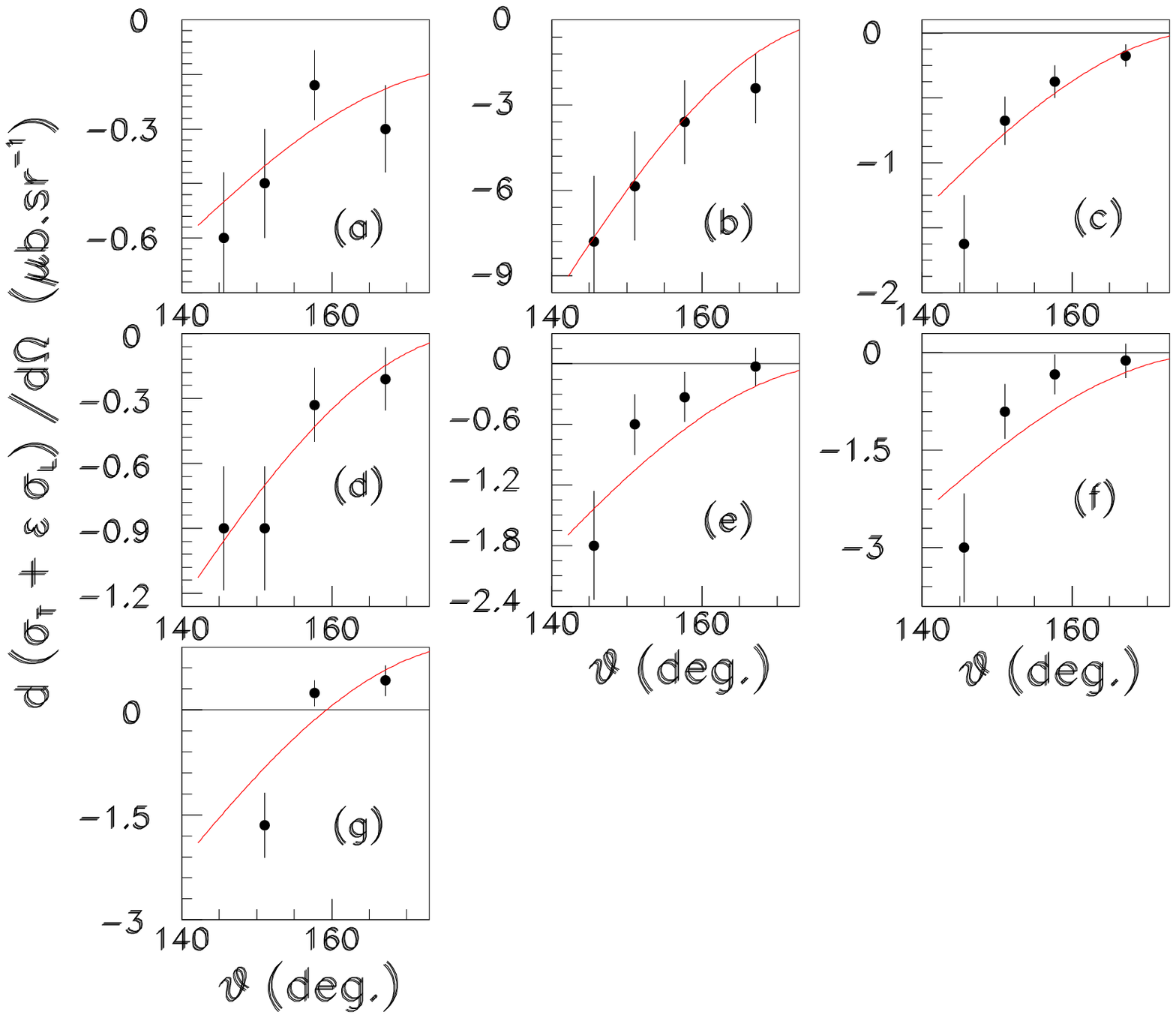}}
\caption [Fig.~7] {Angular variations of the $\sigma_{T}~+~\epsilon\sigma_{L}$ structure function
for the seven narrow structure masses
as extracted from Figs.~1 and 2.  (see text).  Inserts (a), (b), (c), (d), (e), (f),  and (g) correspond
respectively to the following masses: M=1136~MeV, 1210~MeV,  1277~MeV, 1339~MeV, 1384~MeV, 1480~MeV, and 1540~MeV.}
\end{figure}
\end{center}

\begin{center}
\begin{figure}[!ht]
\scalebox{0.9}[0.9]{
\includegraphics[bb=12 12  530 560,clip,scale=1]{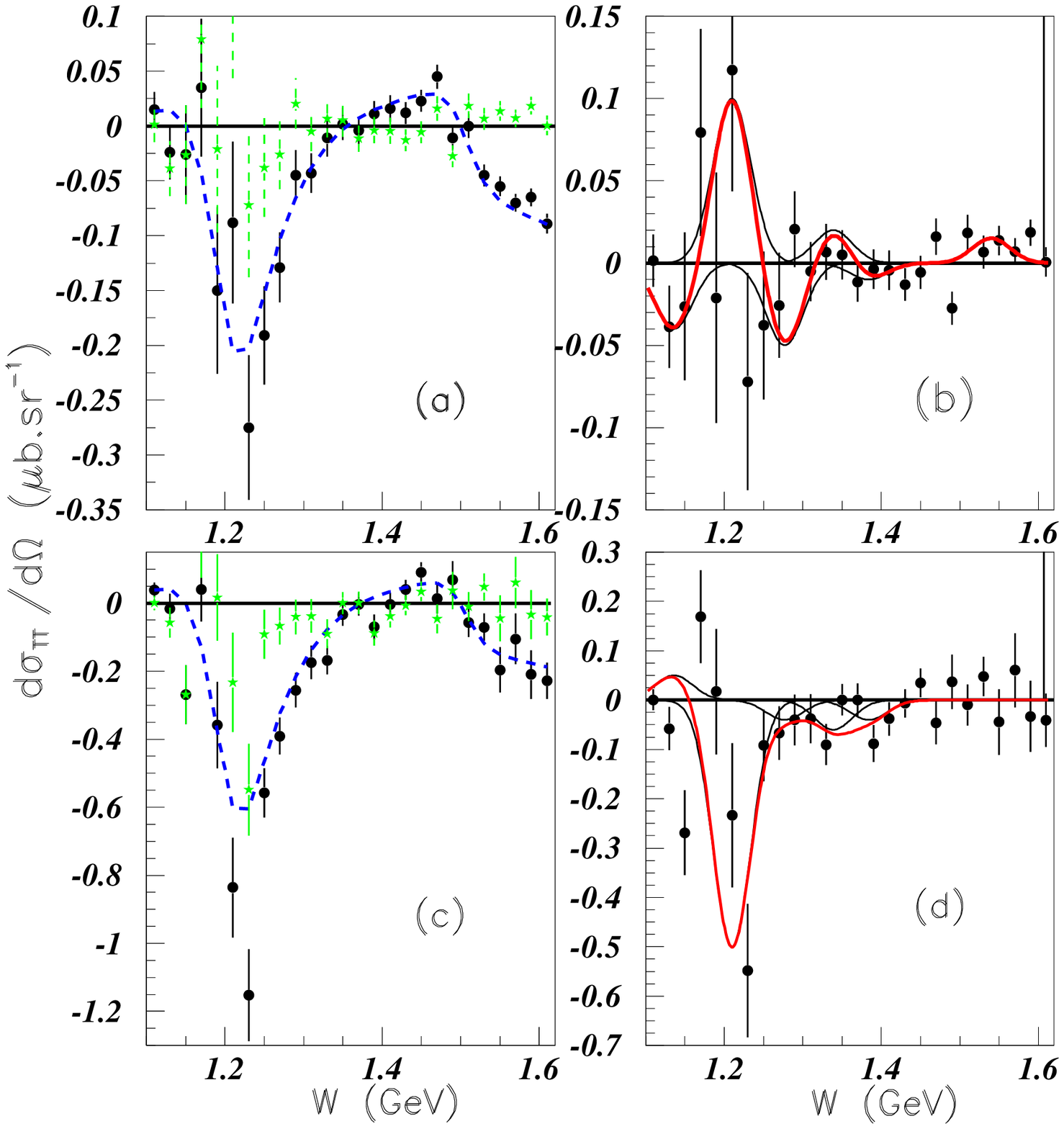}}
\caption [Fig.~8] {Same  as in Fig.~5 but for $\sigma_{TT}$ structure function.}
\end{figure}
\end{center}

\newpage\clearpage

\begin{center}
\begin{figure}[!ht]
\scalebox{0.9}[0.9]{
\includegraphics[bb=12 12  530 560,clip,scale=1]{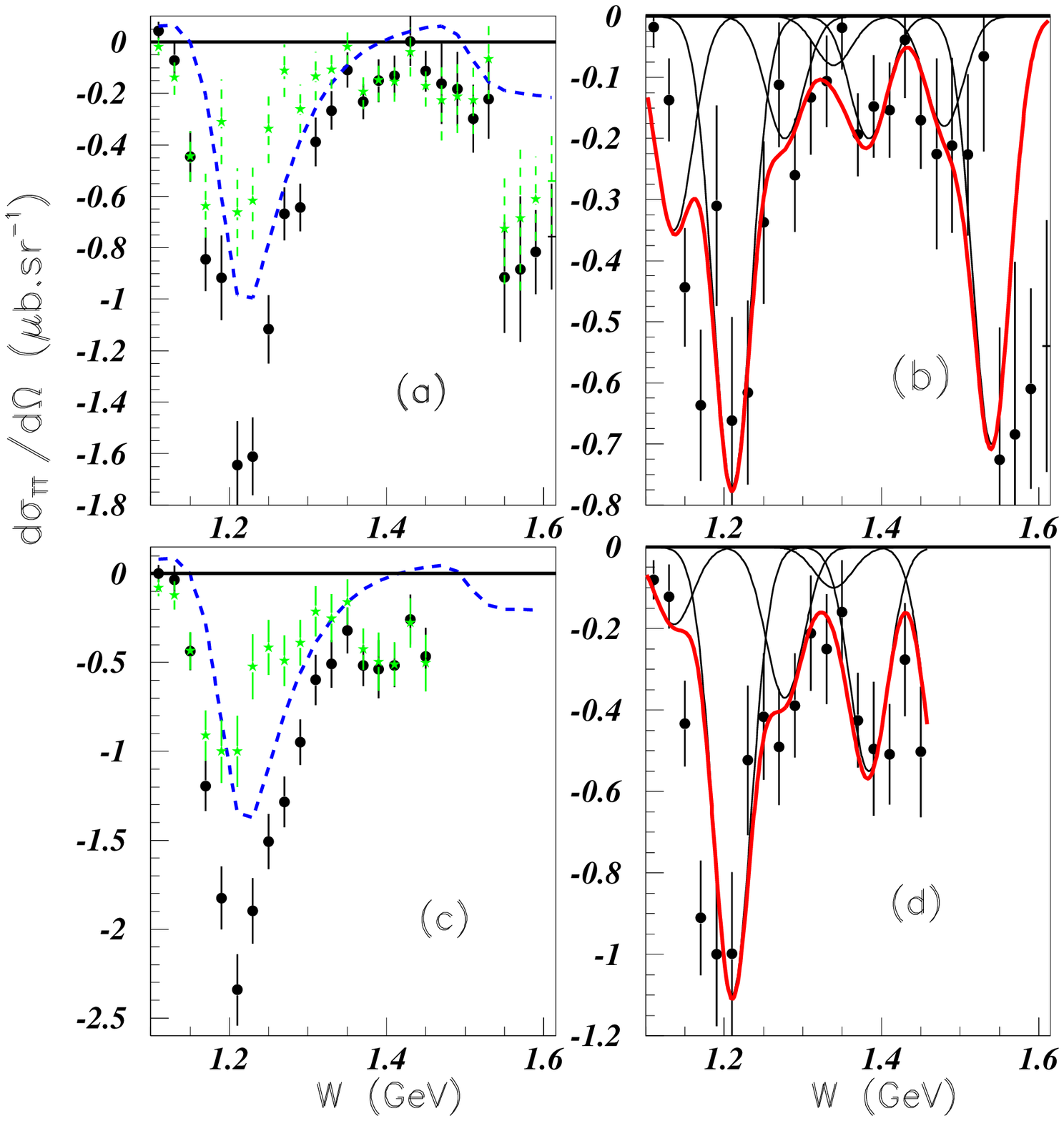}}
\caption [Fig.~9] {Same  as in Fig.~6 but for $\sigma_{TT}$ structure function.}
\end{figure}
\end{center}

\newpage\clearpage
\begin{center}
\begin{figure}[!ht]
\scalebox{0.9}[0.9]{
\includegraphics[bb=12 12  530 450,clip,scale=1]{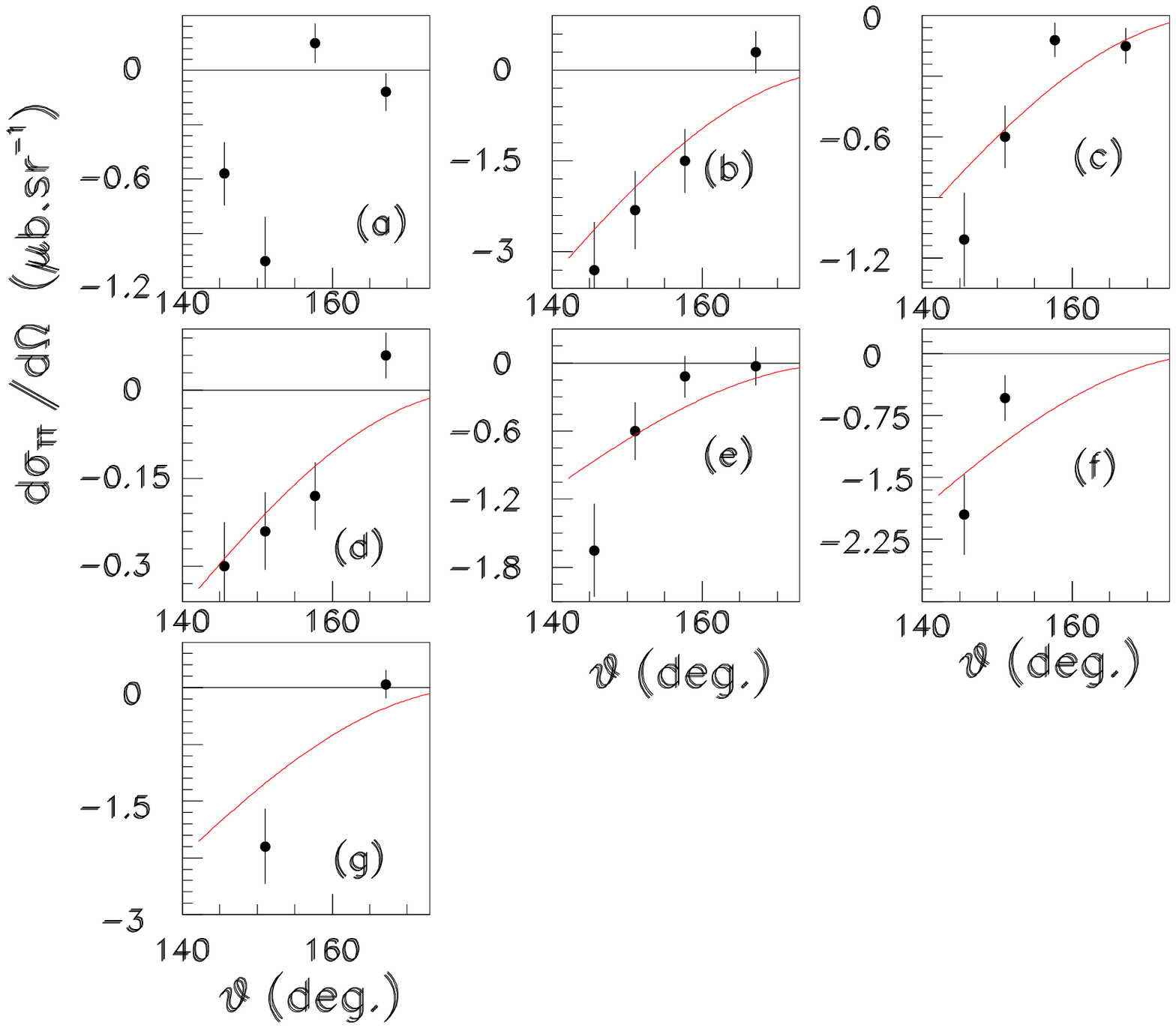}}
\caption [Fig.~10] {Angular variations of  $\sigma_{TT}$ structure function for the seven
 narrow structure masses as extracted from Figs.~8 and 9  (see text).  Inserts (a), (b), (c),
 (d), (e), (f), and (g) correspond respectively to the following masses: 
M=1136~MeV, 1210~MeV, 1277~MeV, 
1339~MeV, 1384~MeV, 1480~MeV, and 1540~MeV.}
\end{figure}
\end{center}

\begin{center}
\begin{figure}[!ht]
\scalebox{0.9}[0.9]{
\includegraphics[bb=12 12  530 560,clip,scale=1]{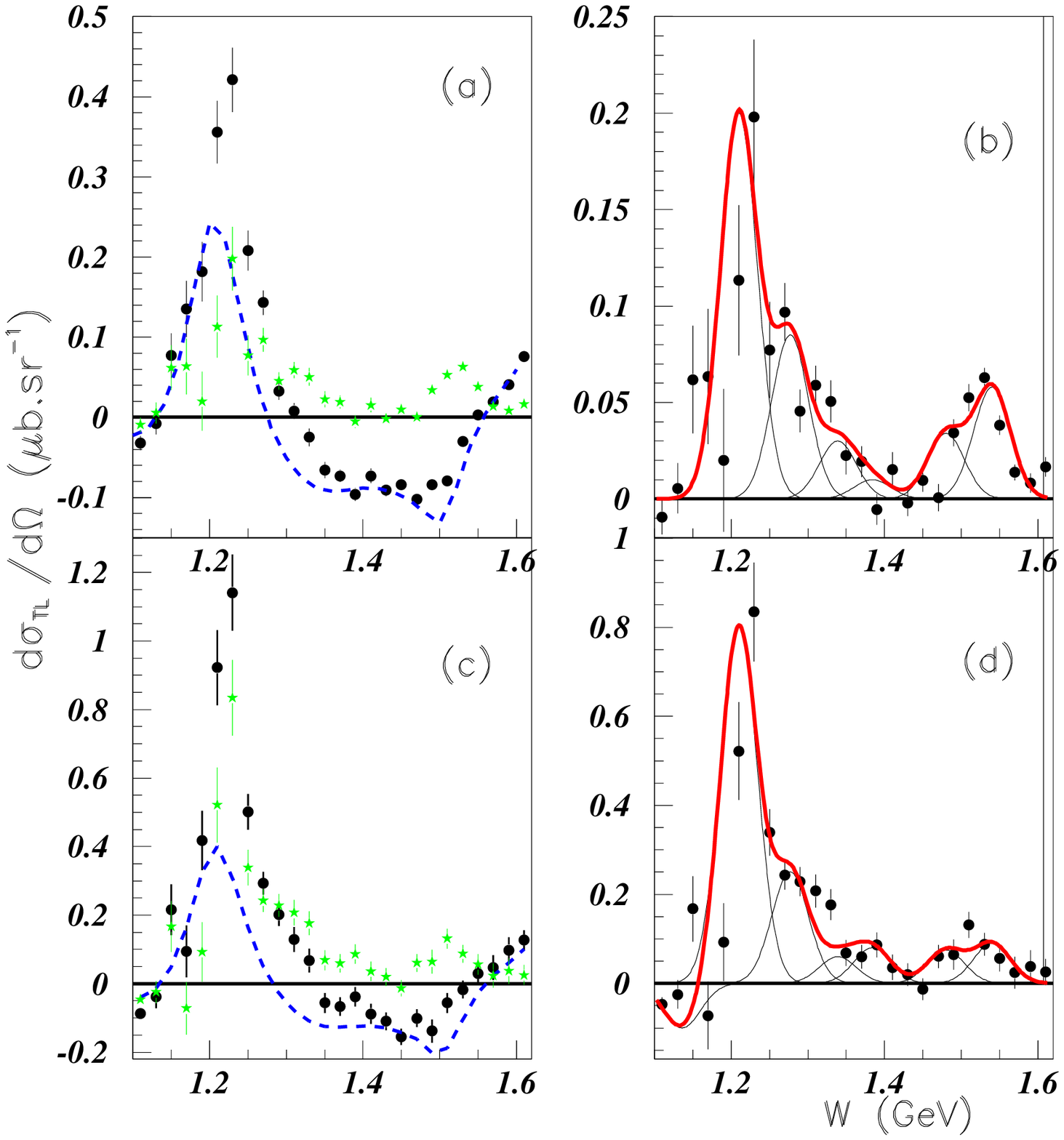}}
\caption [Fig.~11] {Same  as in Fig.~5 but for $\sigma_{TL}$ structure function.}
\end{figure}
\end{center}

\newpage\clearpage
\begin{center}
\begin{figure}[!ht]
\scalebox{0.9}[0.9]{
\includegraphics[bb=12 12  530 560,clip,scale=1]{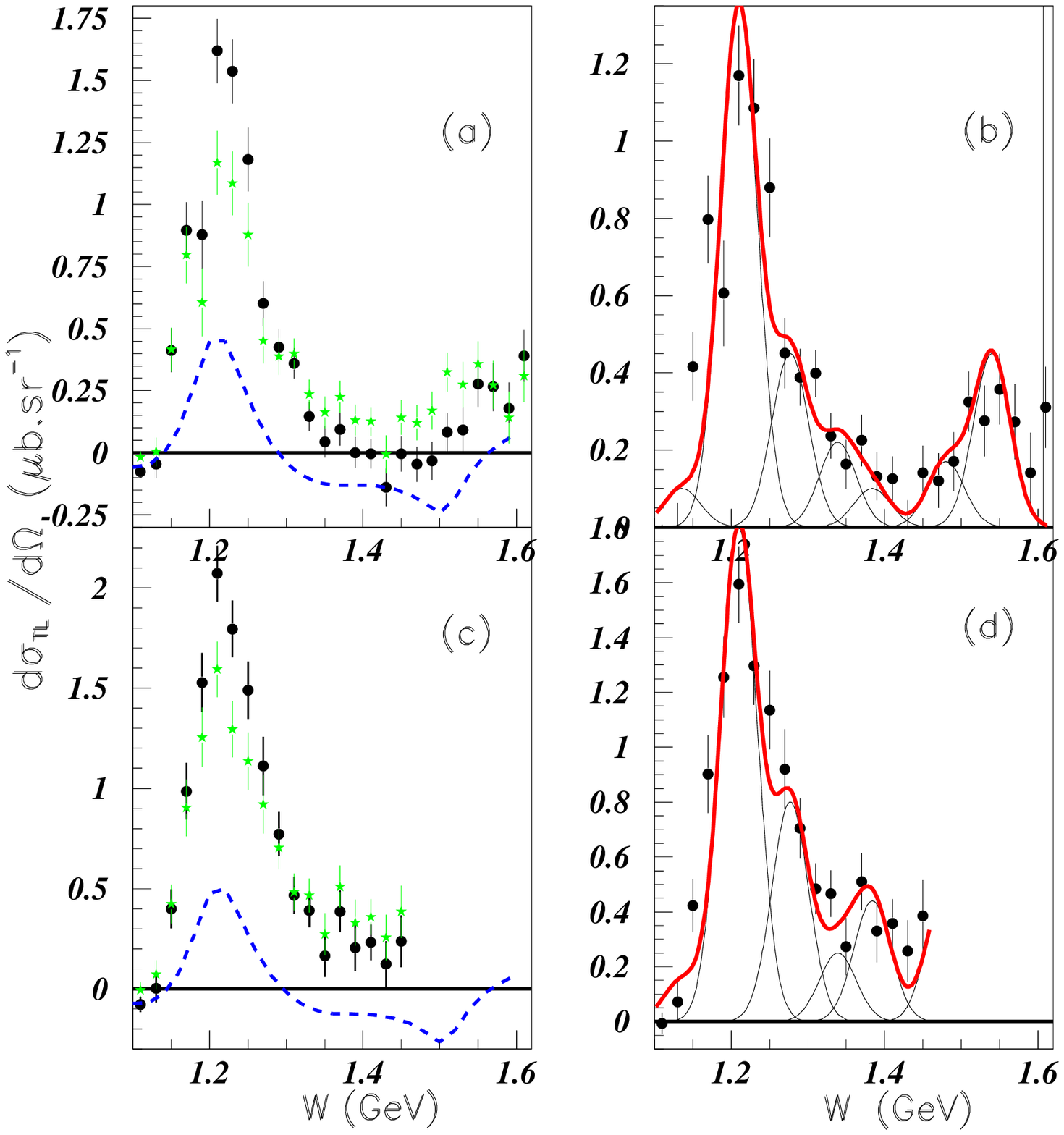}}
\caption [Fig.~12] {Same  as in Fig.~6 but for $\sigma_{TL}$ structure function.}
\end{figure}
\end{center}

\begin{center}
\begin{figure}[!ht]
\scalebox{0.9}[0.9]{
\includegraphics[bb=12 12  530 450,clip,scale=1]{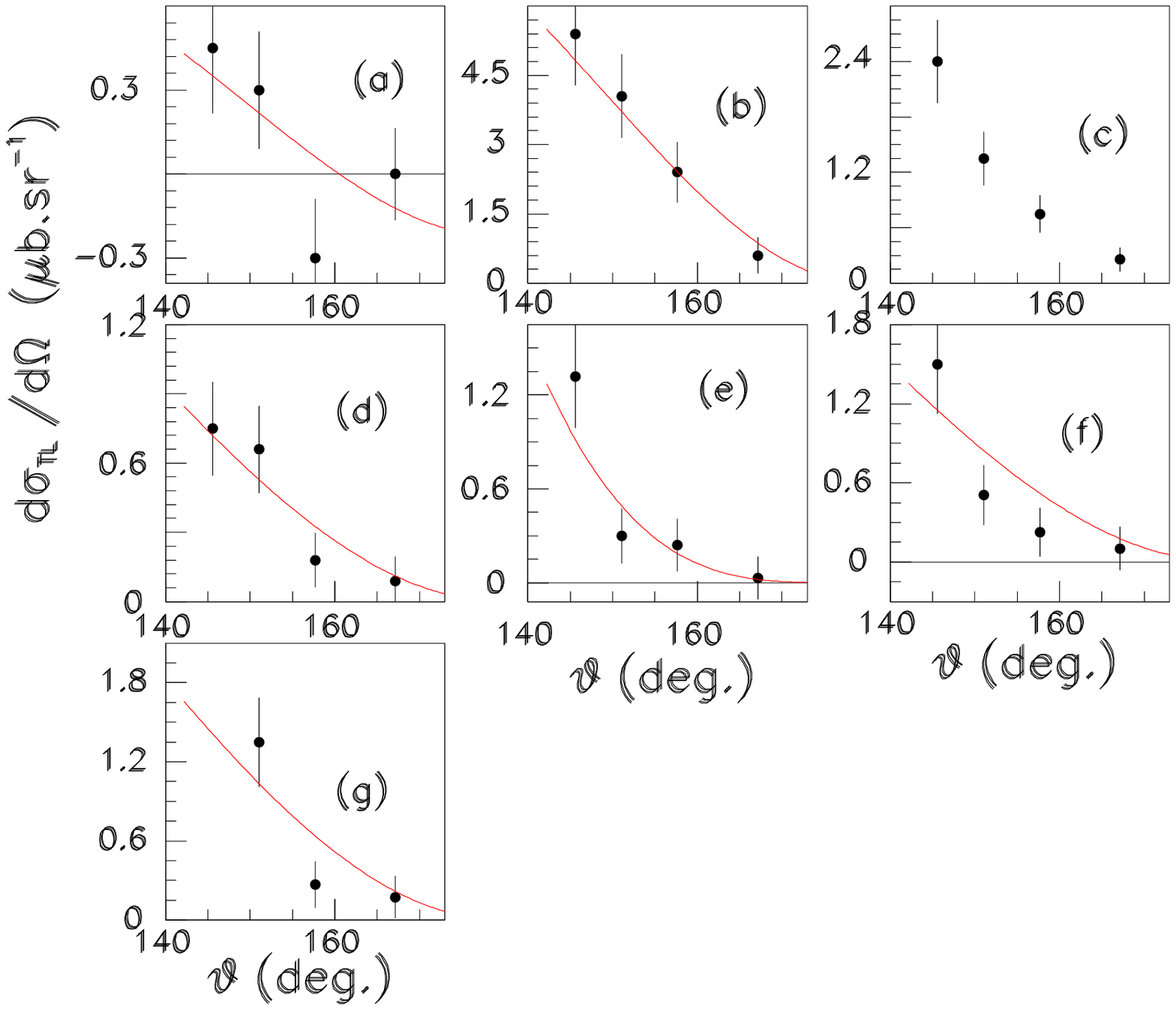}}
\caption [Fig.~13] {Same  as in Fig.~7, but for $\sigma_{TL}$ structure function.}
\end{figure}
\end{center}

\newpage\clearpage
\begin{center}

\begin{figure}[!ht]
\scalebox{0.9}[0.9]{
\includegraphics[bb=12 12  530 550,clip,scale=1]{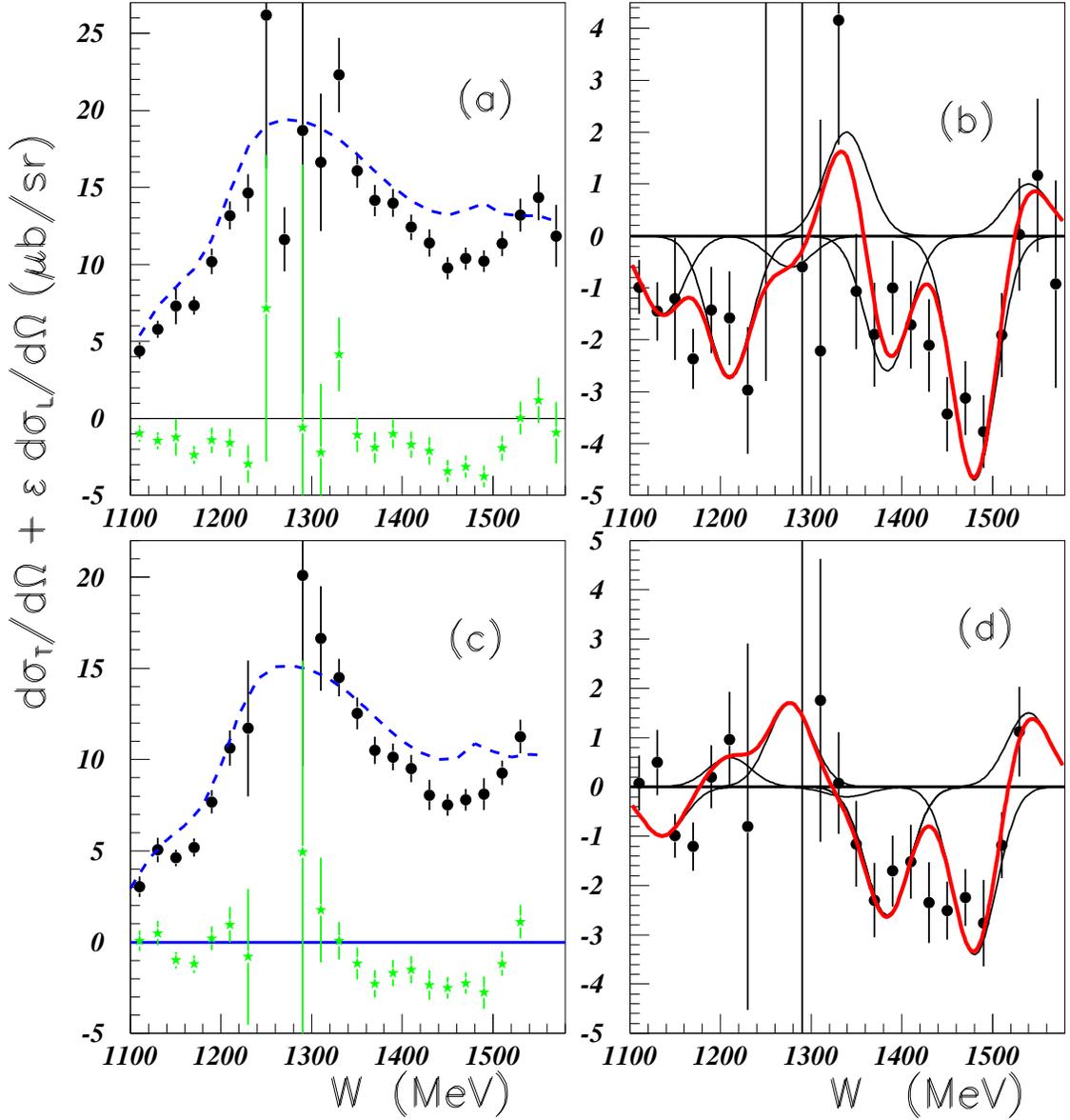}}
\caption [Fig.~14] {$\sigma_{T}~+~\epsilon\sigma_{L}$ structure function of the
$\gamma^{*}$p$\to\pi^{+}$n reaction \protect\cite{egiy}, at $\theta$=7.5$^{0}$
 and Q$^{2}$=0.3~GeV$^{2}$ in inserts (a) and (b) and at $\theta$=7.5$^{0}$ and 
Q$^{2}$=0.4~GeV$^{2}$ in inserts (c) and (d) (see text).}
\end{figure}
\end{center}
\newpage\clearpage

\begin{center}
\begin{figure}[!ht]
\scalebox{0.9}[0.9]{
\includegraphics[bb=12 12  530 550,clip,scale=1]{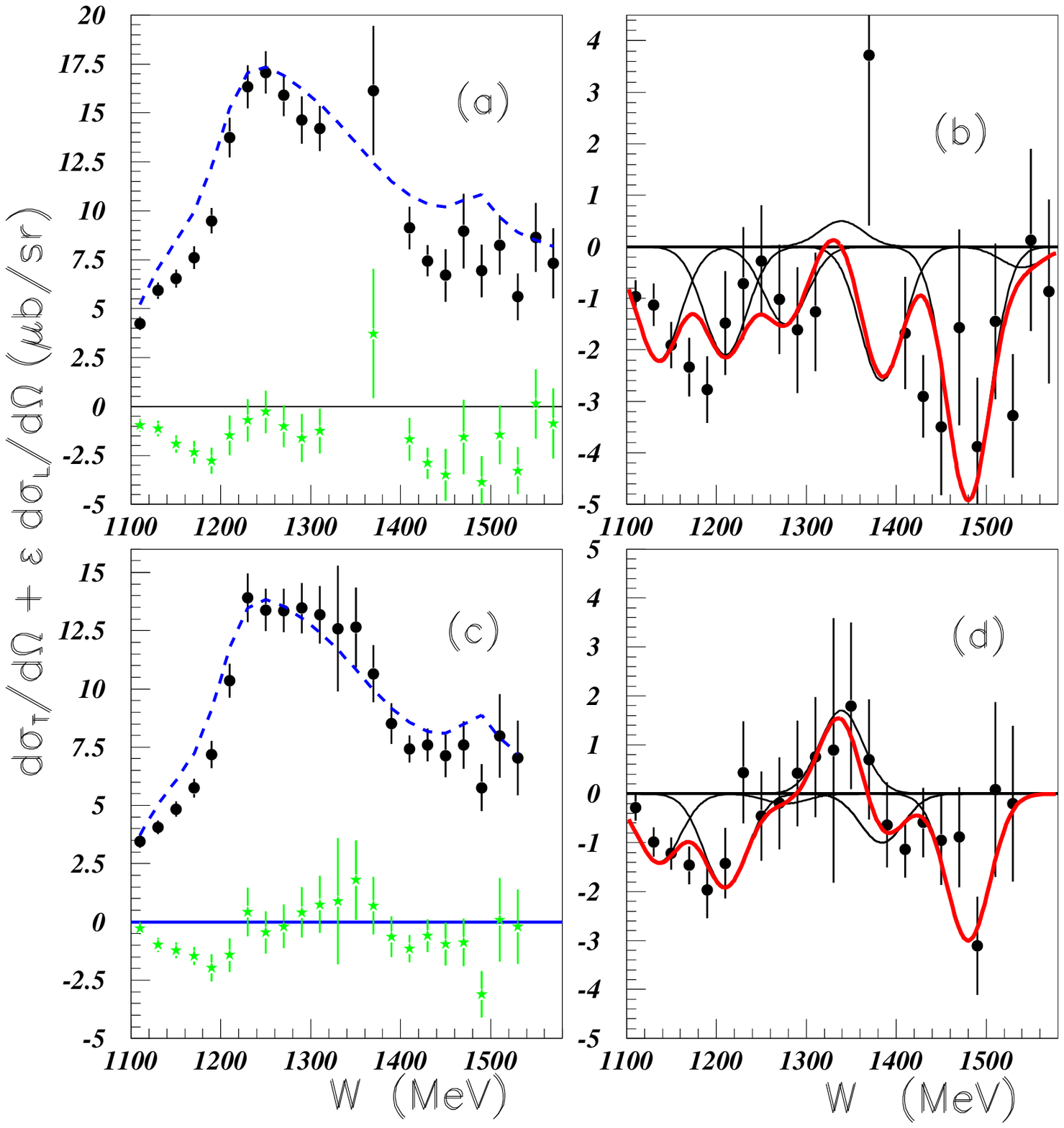}}
\caption [Fig.~15] {Same  as in Fig.~14 but for $\theta$=22.5$^{0}$ .}
\end{figure}
\end{center}

\newpage\clearpage
\begin{center}
\begin{figure}[!ht]
\scalebox{0.9}[0.9]{
\includegraphics[bb=12 12  530 550,clip,scale=1]{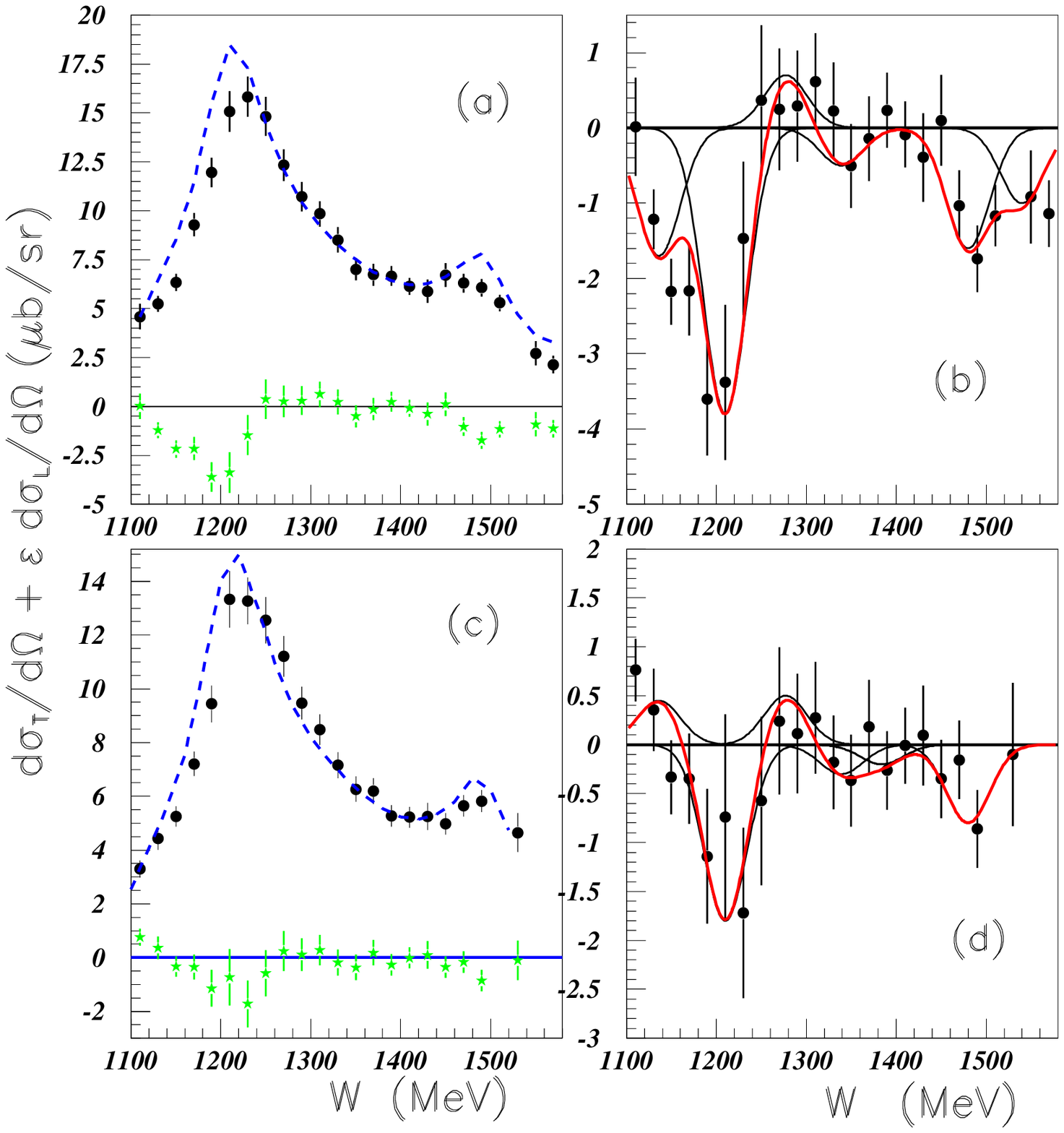}}
\caption [Fig.~16] {Same  as in Fig.~14 but for $\theta$=52.5$^{0}$ .}
\end{figure}
\end{center}

\begin{center}
\begin{figure}[!ht]
\scalebox{0.9}[0.9]{
\includegraphics[bb=12 12  530 550,clip,scale=1]{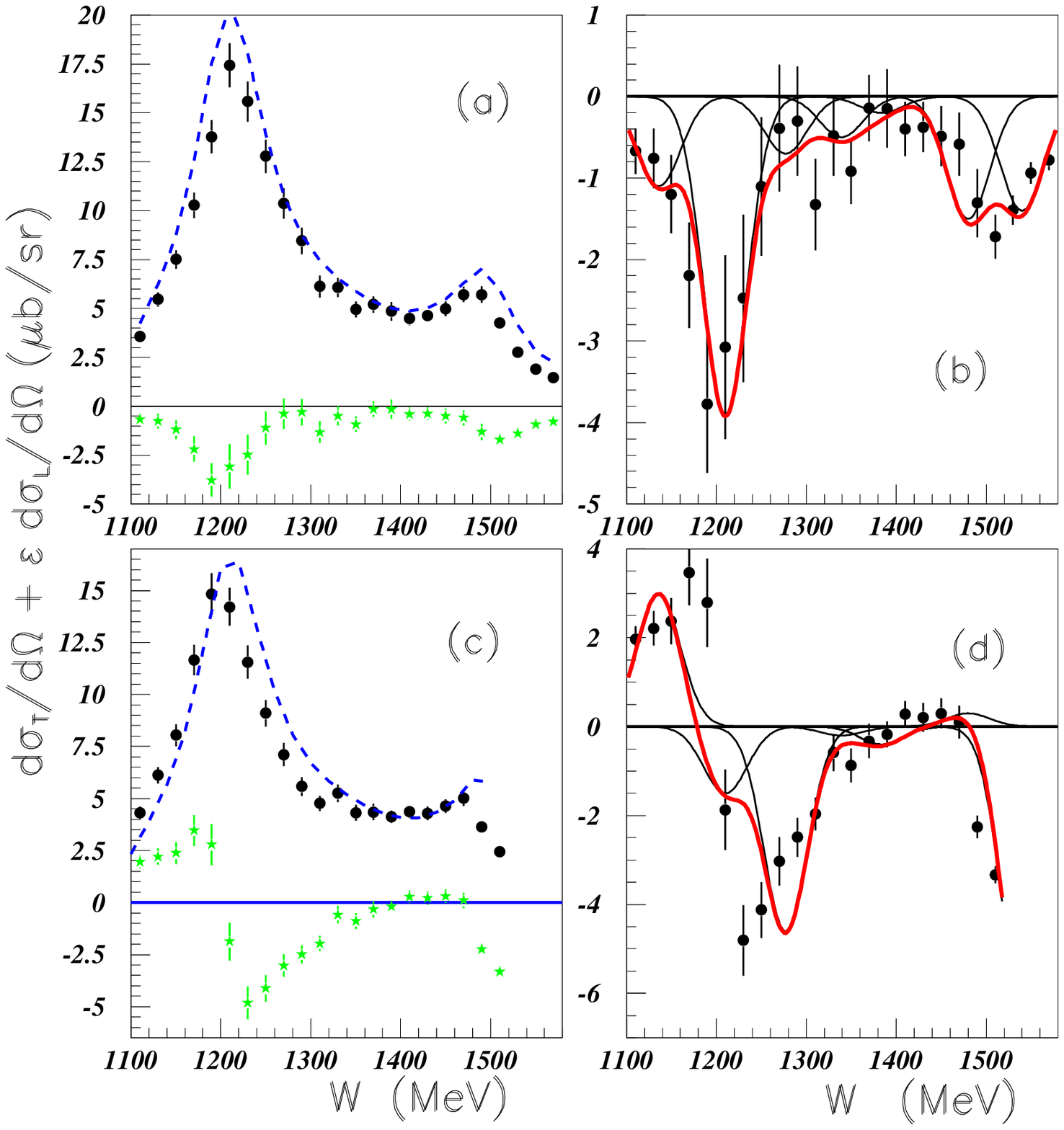}}
\caption [Fig.~17] {Same  as in Fig.~14 but for $\theta$=67.5$^{0}$ .}
\end{figure}
\end{center}

\newpage\clearpage
\begin{center}
\begin{figure}[!ht]
\scalebox{0.9}[0.9]{
\includegraphics[bb=12 12  530 550,clip,scale=1]{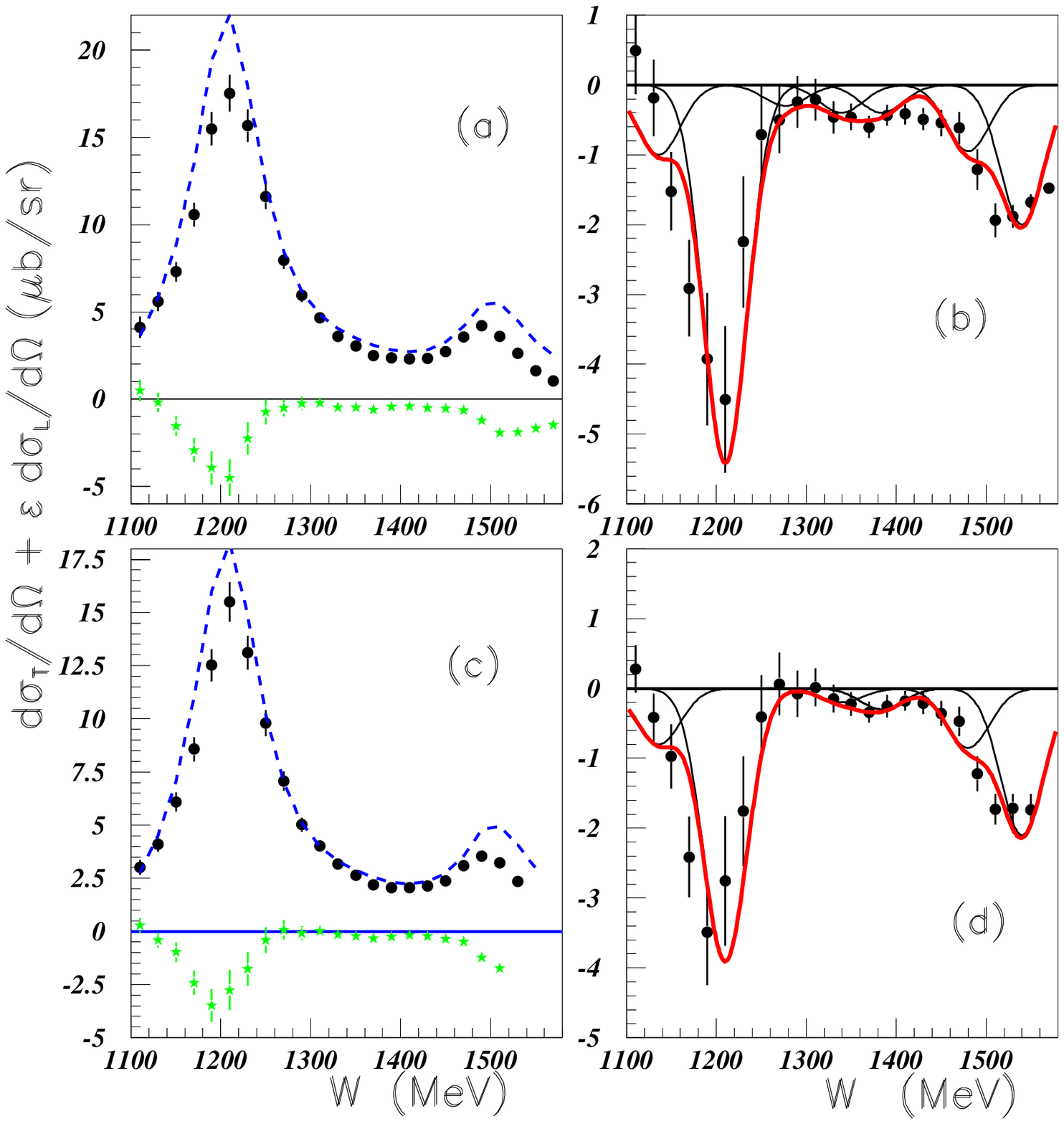}}
\caption [Fig.~18] {Same  as in Fig.~14 but for $\theta$=97.5$^{0}$ .}
\end{figure}
\end{center}

\newpage\clearpage
\begin{center}
\begin{figure}[!ht]
\scalebox{0.9}[0.9]{
\includegraphics[bb=12 12  530 550,clip,scale=1]{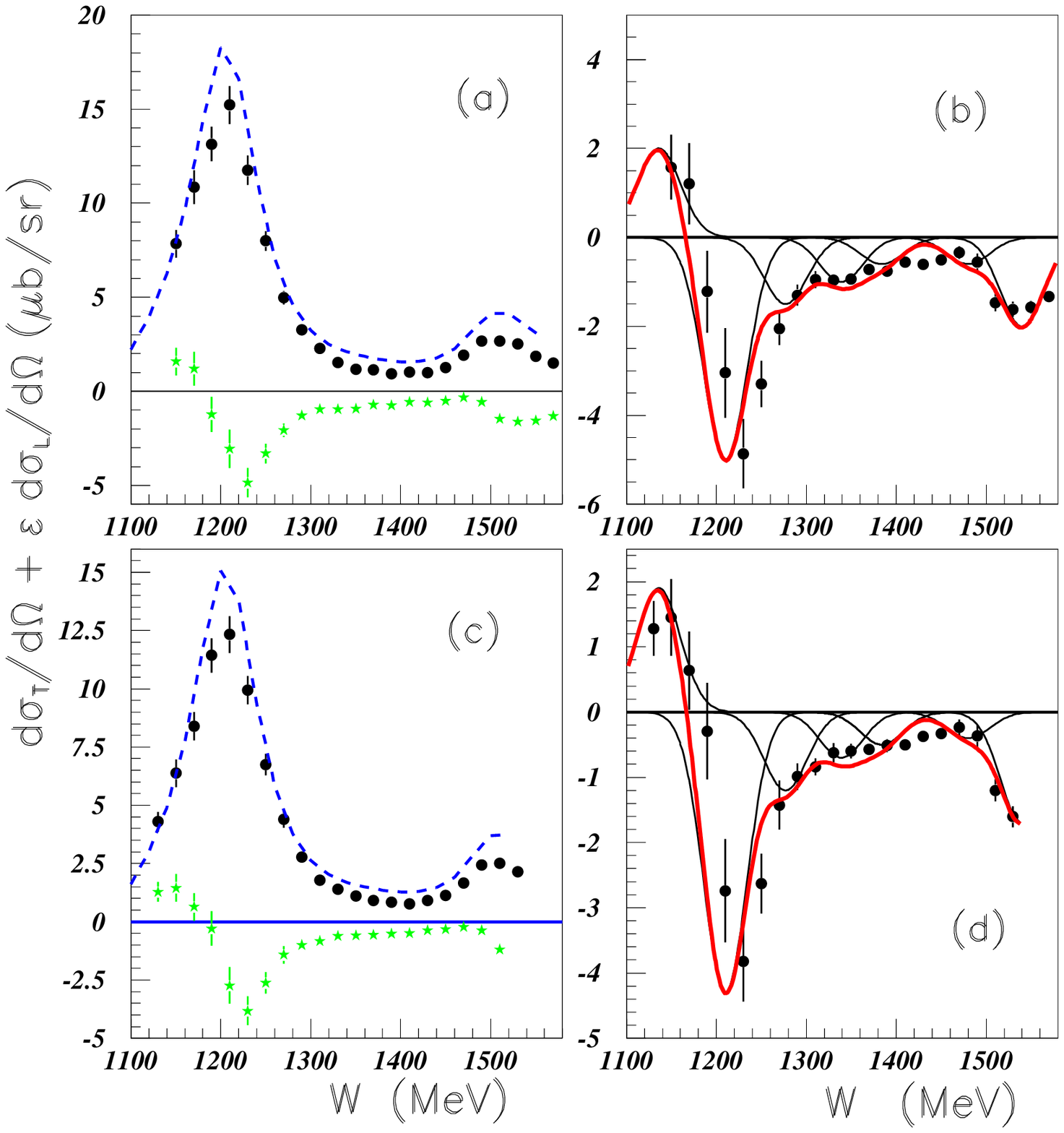}}
\caption [Fig.~19] {Same  as in Fig.~14 but for $\theta$=127.5$^{0}$ .}
\end{figure}
\end{center}

\begin{center}
\begin{figure}[!ht]
\scalebox{0.9}[0.9]{
\includegraphics[bb=42 12  530 550,clip,scale=1]{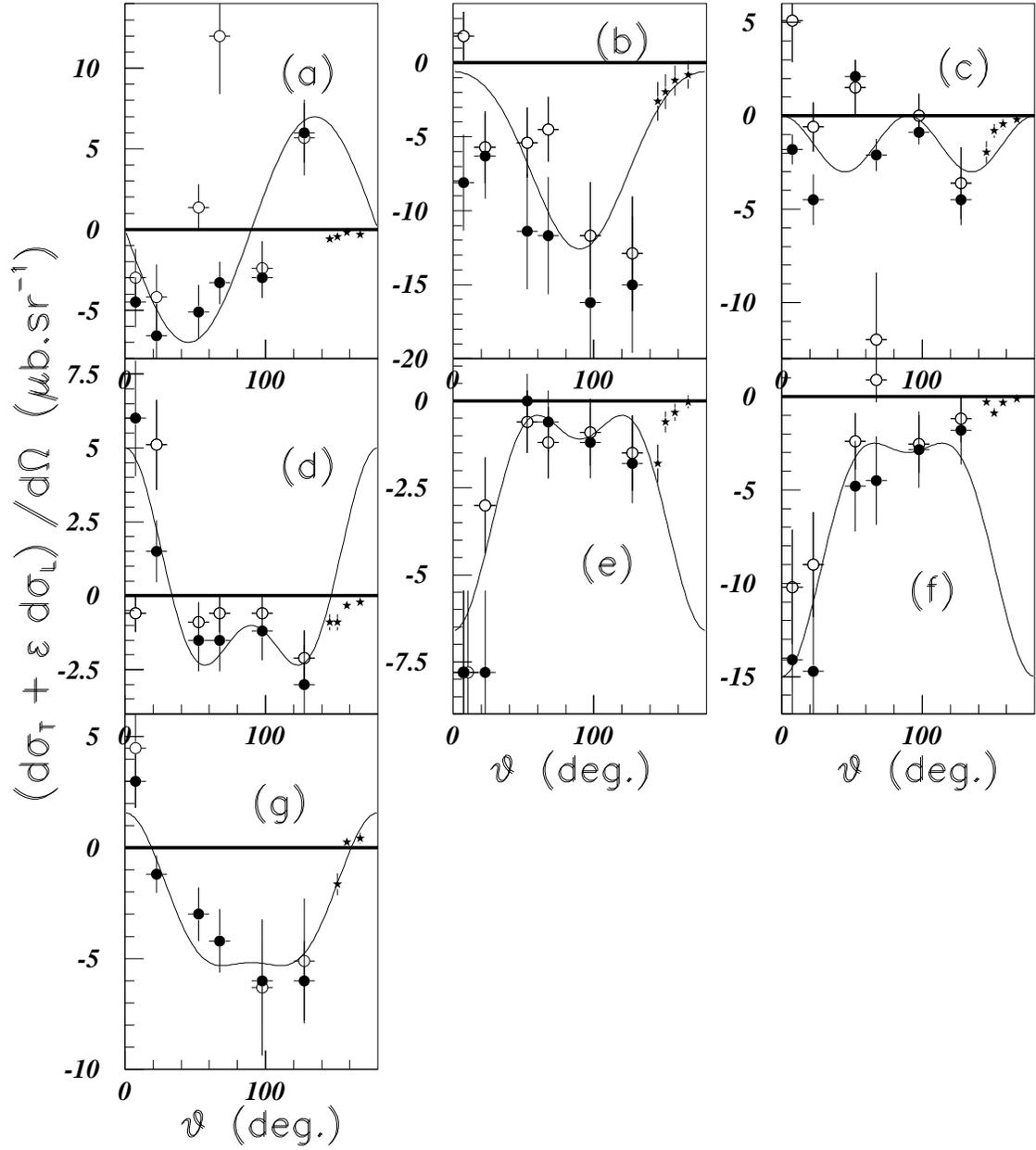}}
\caption [Fig.~20] {Same  as in Fig.~7 but showing the angular
variation of the yield of the 
$\sigma_{T}~+~\epsilon\sigma_{L}$ structure function for both reactions.
Full circles correspond to Q$^{2}$=0.3~GeV$^{2}$, empty circles correspond to Q$^{2}$=0.4~GeV$^{2}$, and stars correspond to Q$^{2}$=1~GeV$^{2}$ (see text).}
\end{figure}
\end{center}

\begin{center}
\begin{figure}[t]
\suppressfloats[!]
\scalebox{0.9}[0.9]{
\includegraphics[bb=12 12  530 550,clip,scale=1]{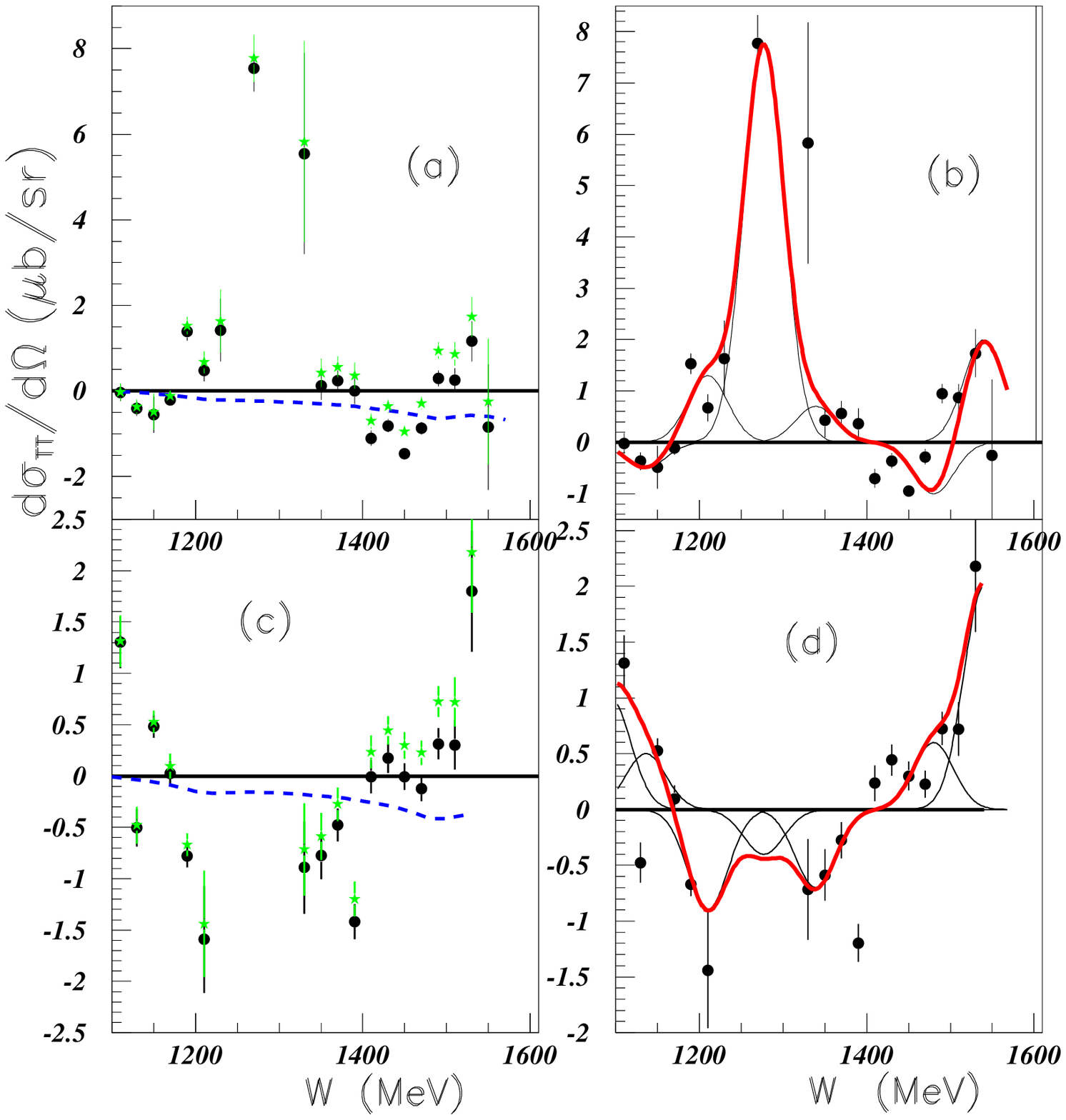}}
\caption [Fig.~21] {Cross-section of the $\sigma_{TT}$ structure function 
of the ep$\to$e'n$\pi^{+}$ reaction at $\theta$=7.5$^{0}$. Inserts (a) and (b) show
the results at Q$^{2}$=0.3~GeV$^{2}$, inserts (c) and (d) show the results at 
Q$^{2}$=0.4~GeV$^{2}$. }
\end{figure}
\end{center}

\begin{center}
\begin{figure}[!ht]
\suppressfloats[!]
\scalebox{0.9}[0.9]{
\includegraphics[bb=12 12  530 550,clip,scale=1]{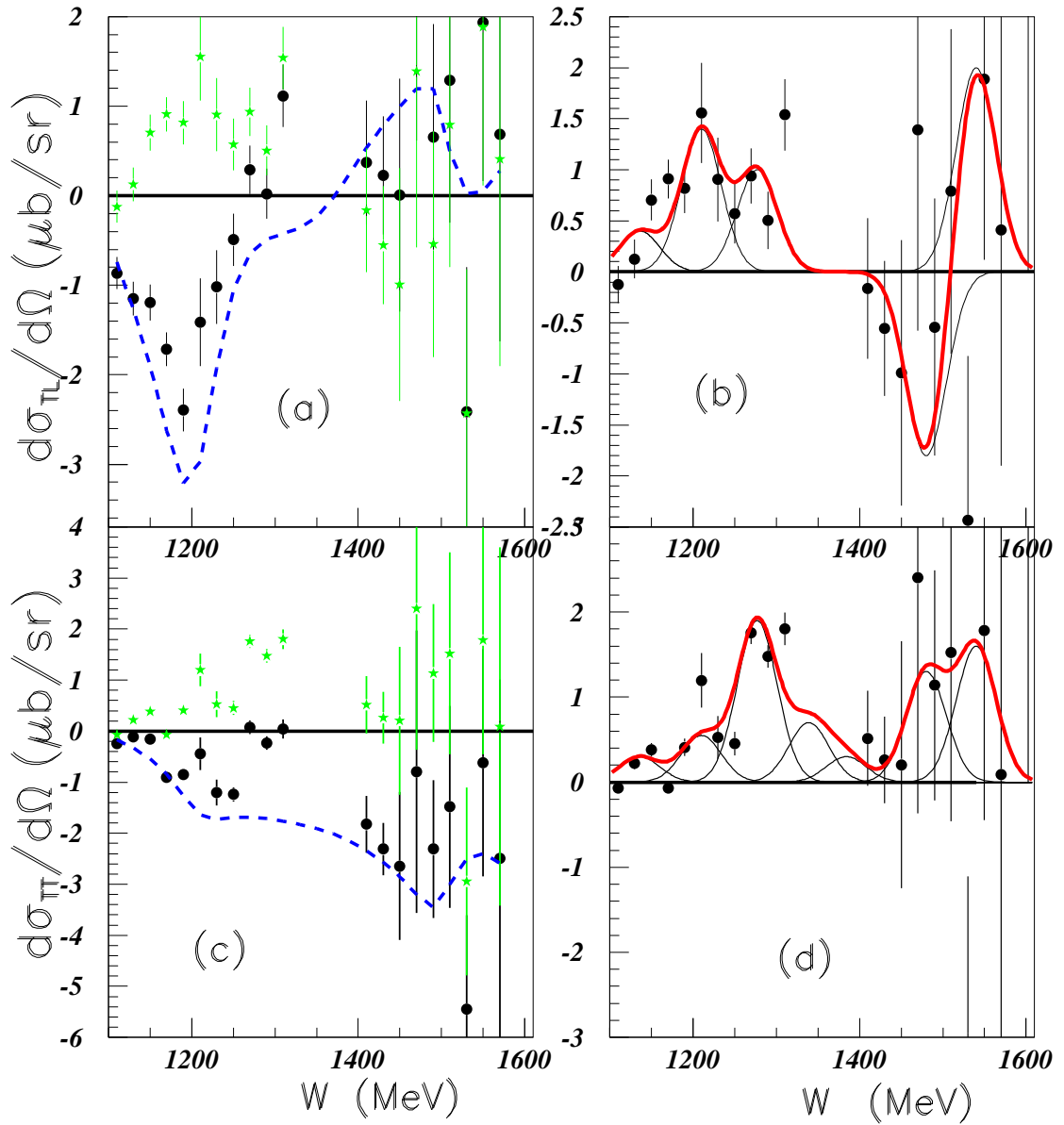}}
\caption [Fig.~22] {Cross-section of two structure functions 
of the ep$\to$e'n$\pi^{+}$ reaction at $\theta$=22.5$^{0}$ and  
Q$^{2}$=0.3~GeV$^{2}$. Inserts (a) and (b) show the $\sigma_{TL}$ 
structure function, and inserts (c) and (d) show the 
 $\sigma_{TT}$ structure function. }
\end{figure}
\end{center}

\begin{center}
\begin{figure}[!ht]
\scalebox{0.9}[0.9]{
\includegraphics[bb=12 12  530 550,clip,scale=1]{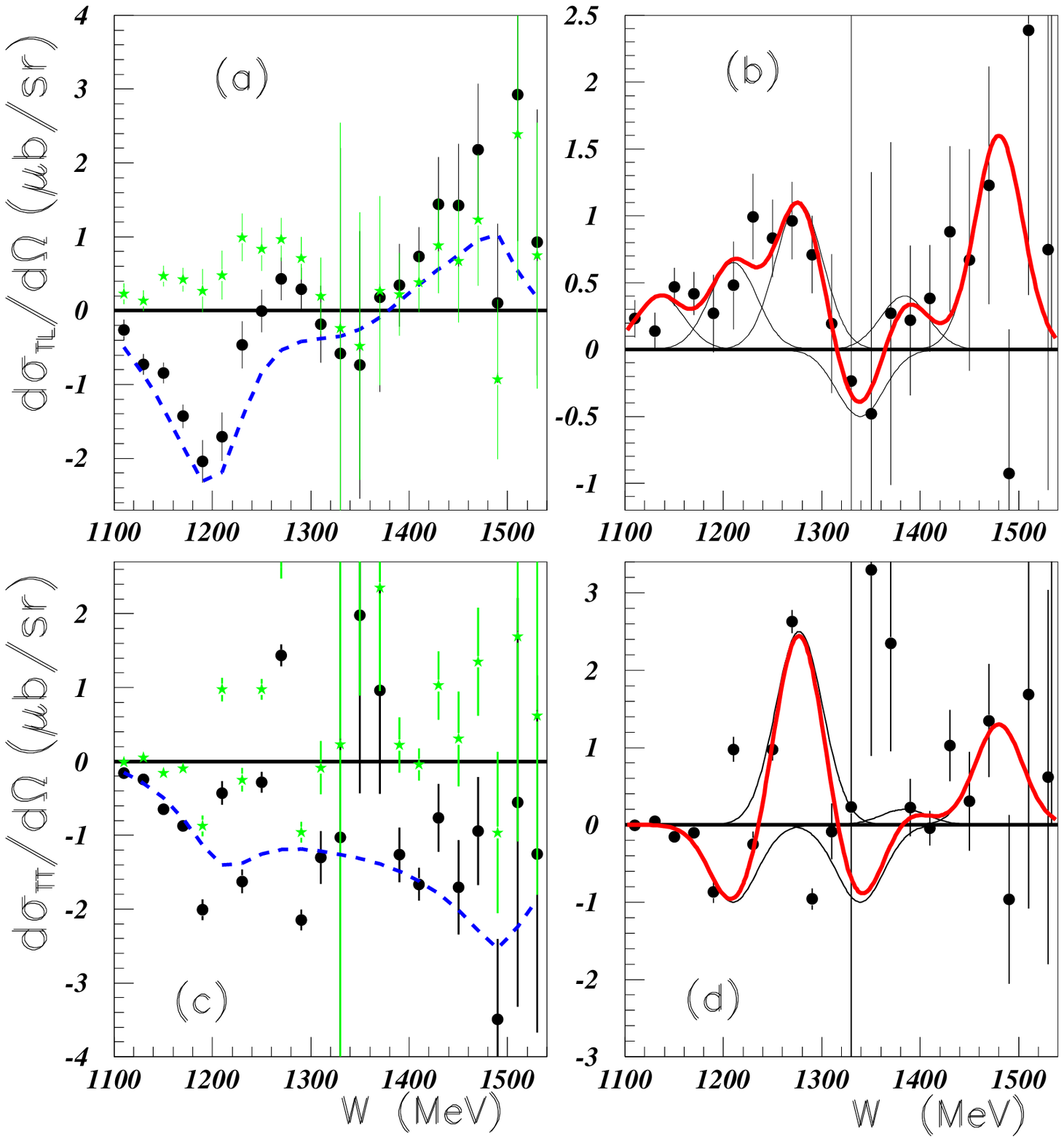}}
\caption [Fig.~23] {Same  as in Fig.~22 but for  Q$^{2}$=0.4~GeV$^{2}$.}
\end{figure}
\end{center}

\begin{center}
\begin{figure}
\scalebox{0.9}[0.9]{
\includegraphics[bb=12 12  530 550,clip,scale=1]{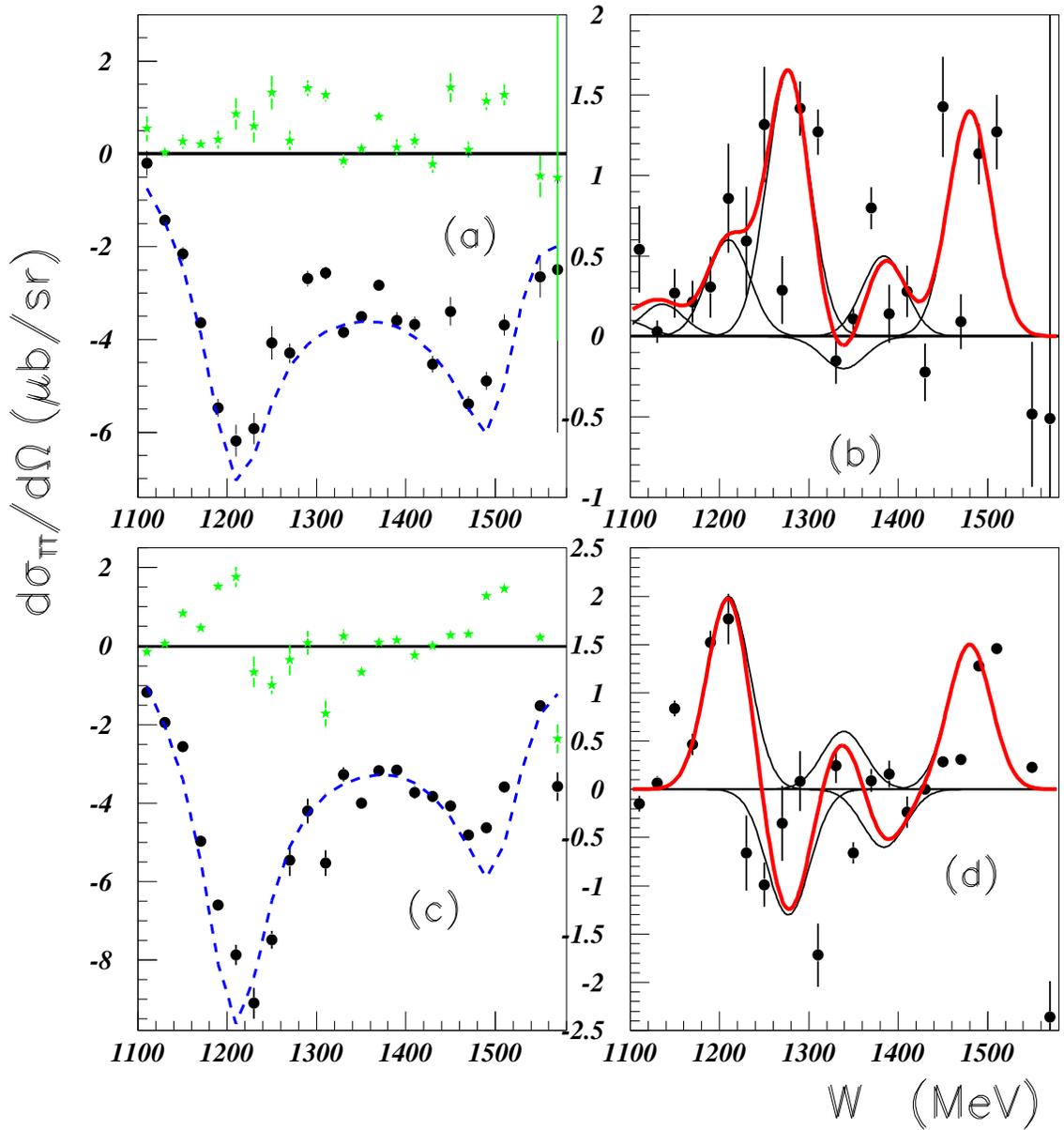}}
\caption [Fig.~24] {Cross-section of the $\sigma_{TT}$ structure function 
of the ep$\to$e'n$\pi^{+}$ reaction at Q$^{2}$=0.3~GeV$^{2}$
$\theta$=52.5$^{0}$ in inserts (a) and (b),  and $\theta$=67.5$^{0}$ in
 inserts (c) and (d).}
\end{figure}
\end{center}

\begin{center}
\begin{figure}[!ht]
\scalebox{0.9}[0.9]{
\includegraphics[bb=12 12  530 550,clip,scale=1]{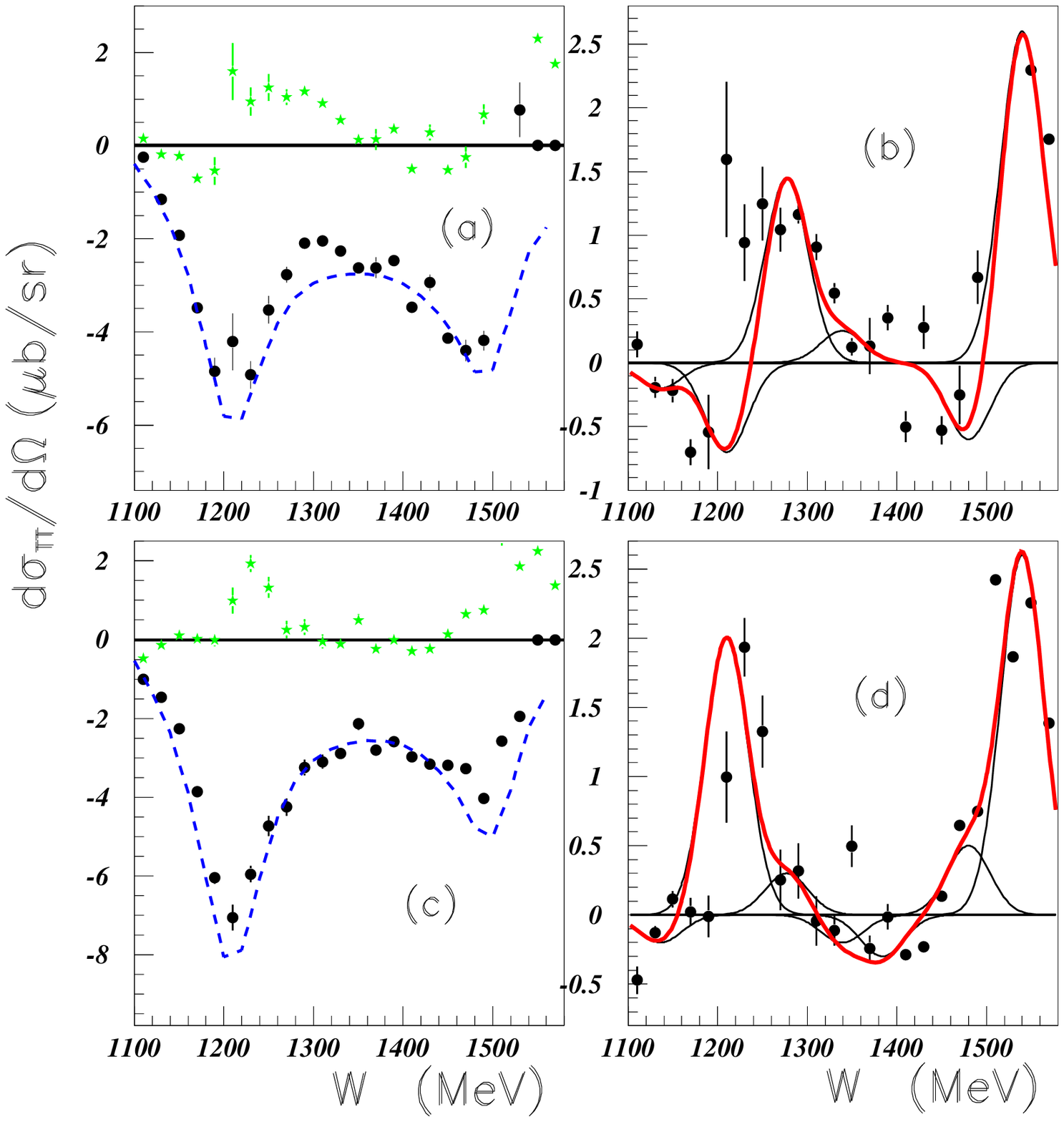}}
\caption [Fig.~25] {Same  as in Fig.~24, but for Q$^{2}$=0.4~GeV$^{2}$.}
\end{figure}
\end{center}

\begin{center}
\begin{figure}[!ht]
\scalebox{0.9}[0.9]{
\includegraphics[bb=12 12  530 550,clip,scale=1]{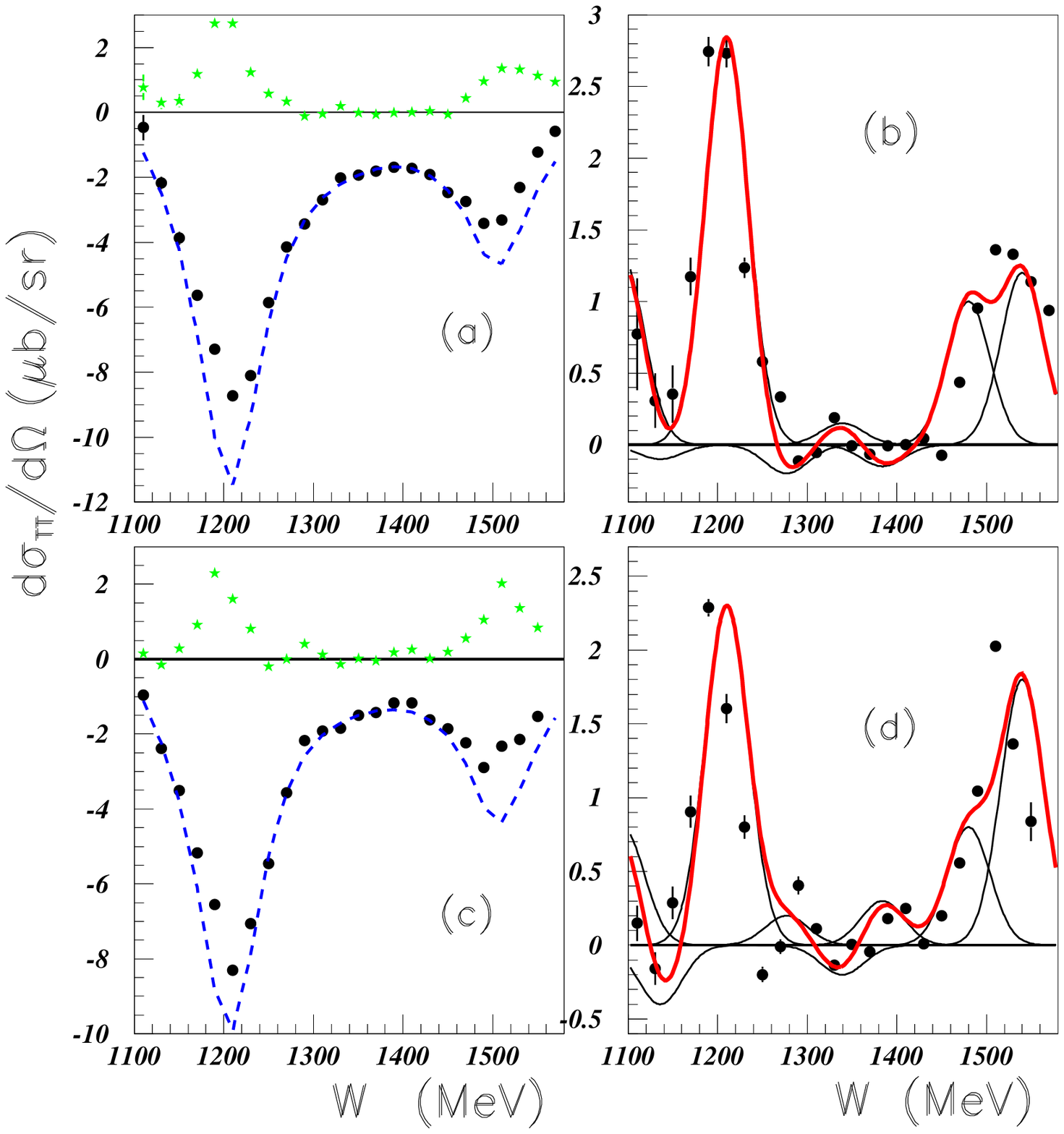}}
\caption [Fig.~26] {Same  as in Fig.~21, but for $\theta$=97.5$^{0}$}
\end{figure}
\end{center}

\begin{center}
\begin{figure}[!ht]
\scalebox{0.9}[0.9]{
\includegraphics[bb=12 12  530 550,clip,scale=1]{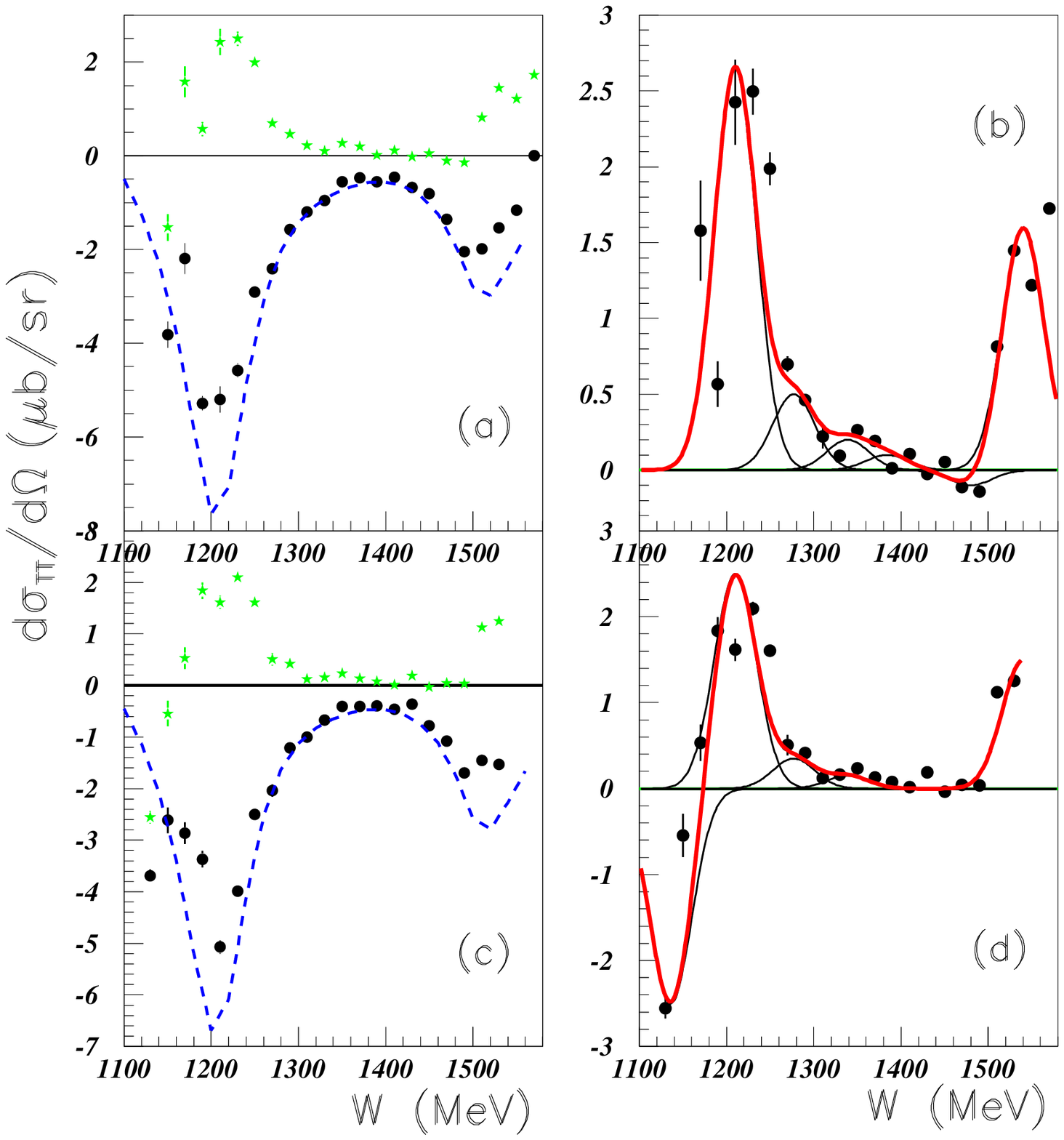}}
\caption [Fig.~27] {Same  as in Fig.~21, but for $\theta$=127.5$^{0}$}
\end{figure}
\end{center}

\begin{center}
\begin{figure}[!ht]
\scalebox{0.9}[0.9]{
\includegraphics[bb=42 12  530 550,clip,scale=1]{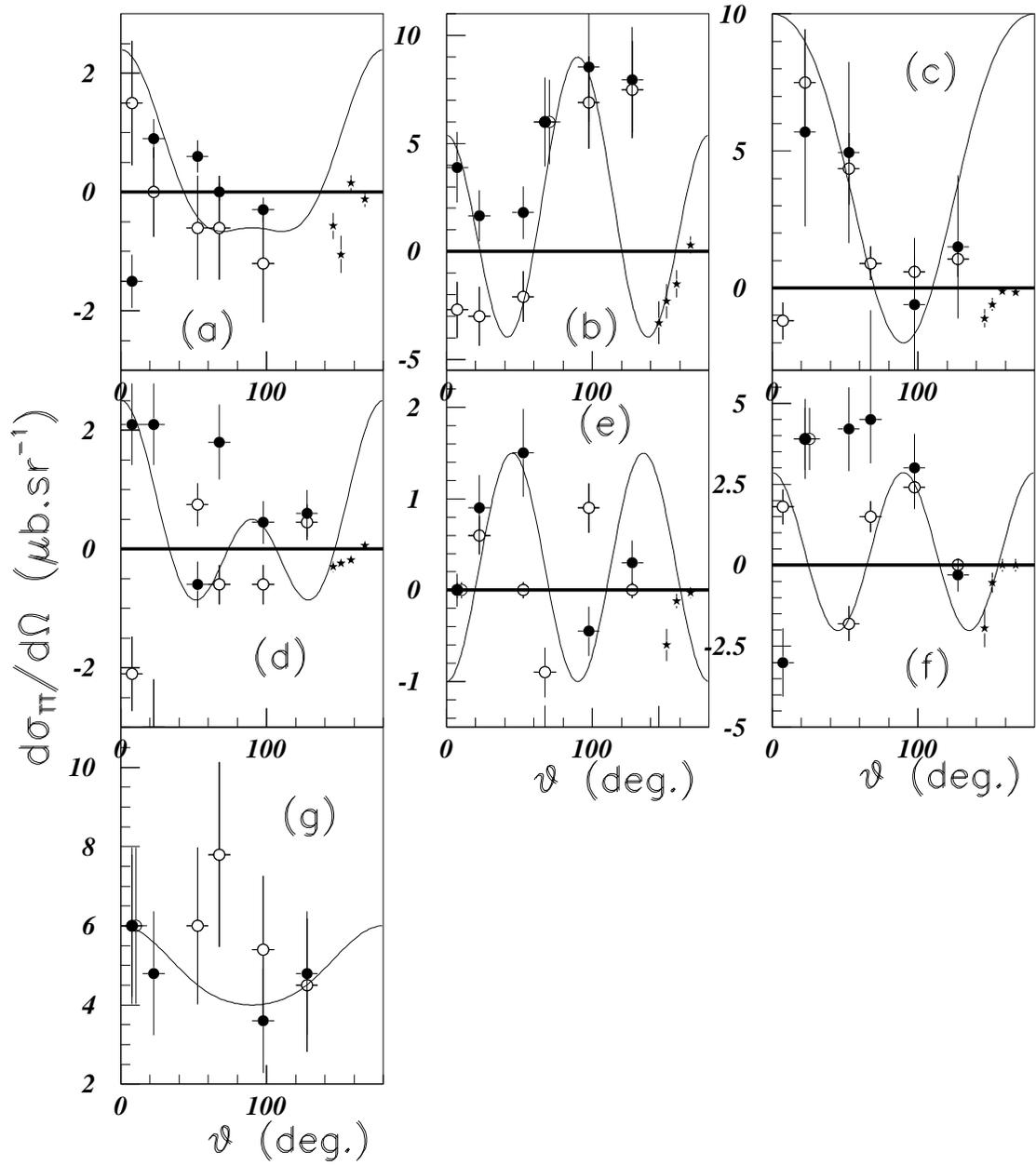}}
\caption [Fig.~28] {Same  as in Fig.~20 but showing the angular
variation of the yield of the $\sigma_{TT}$ structure function for both 
reactions.}
\end{figure}
\end{center}

\newpage\clearpage
\begin{center}
\begin{figure}[!ht]
\scalebox{0.9}[0.9]{
\includegraphics[bb=12 12  530 550,clip,scale=1]{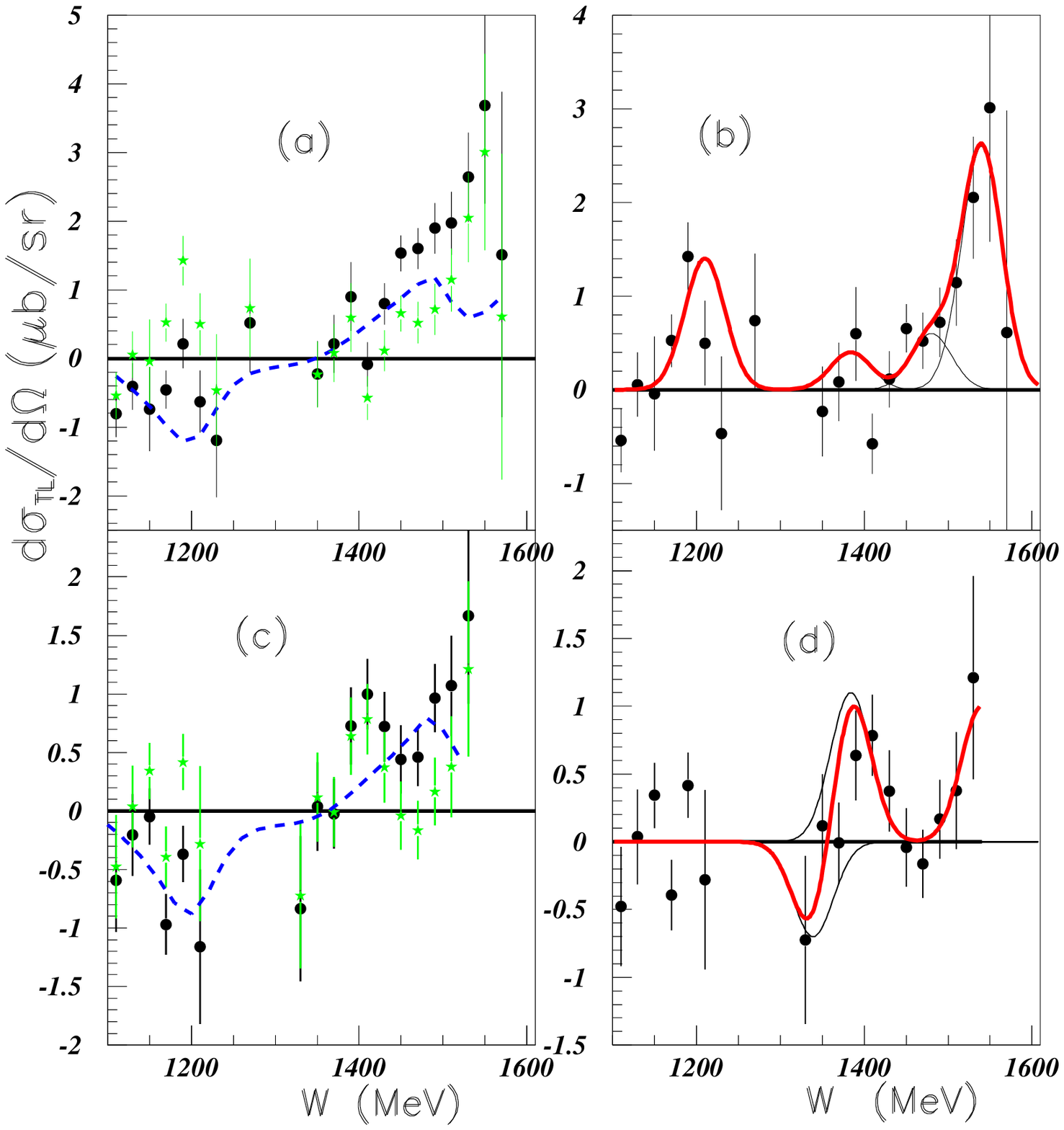}}
\caption [Fig.~29] {Cross-section of the $\sigma_{TL}$ structure function of the 
ep$\to$e'n$\pi^{+}$ reaction at $\theta$=7.5$^{0}$. Inserts (a) and (b) show the
results at Q$^{2}$=0.3~GeV$^{2}$, inserts (c) and (d) show the results at 
 Q$^{2}$=0.4~GeV$^{2}$.}
\end{figure}
\end{center}

\begin{center}
\begin{figure}[!ht]
\scalebox{0.9}[0.9]{
\includegraphics[bb=12 12  530 550,clip,scale=1]{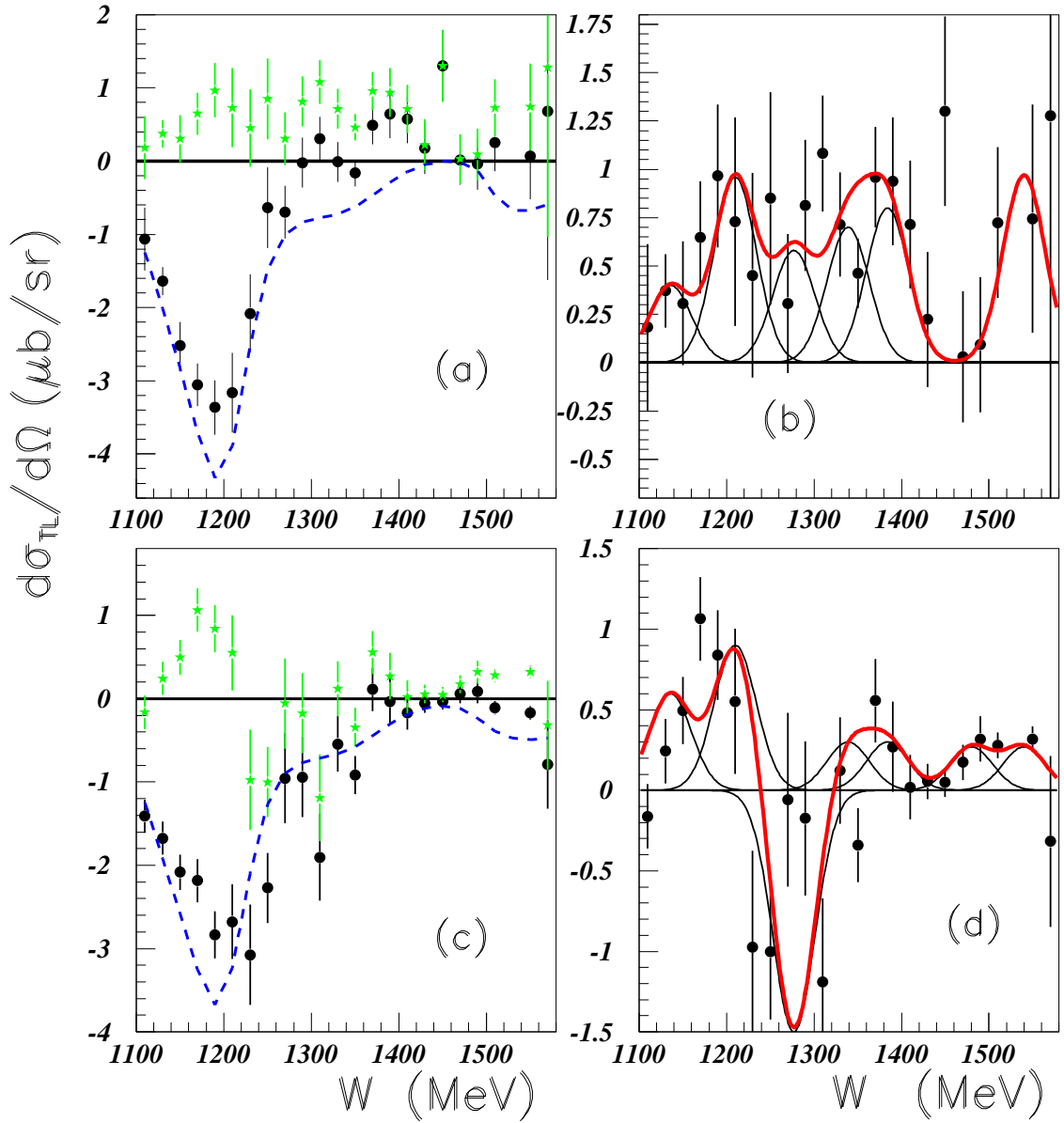}}
\caption [Fig.~30] {Cross-section of the $\sigma_{TL}$ structure function of the 
ep$\to$e'n$\pi^{+}$ reaction at  Q$^{2}$=0.3~GeV$^{2}$.
 Inserts (a) and (b) show the results for for $\theta$=52.5$^{0}$, inserts (c) and (d)
 show the results for $\theta$=67.5$^{0}$.}
\end{figure}
\end{center}

\begin{center}
\begin{figure}[!ht]
\scalebox{0.9}[0.9]{
\includegraphics[bb=12 12  530 550,clip,scale=1]{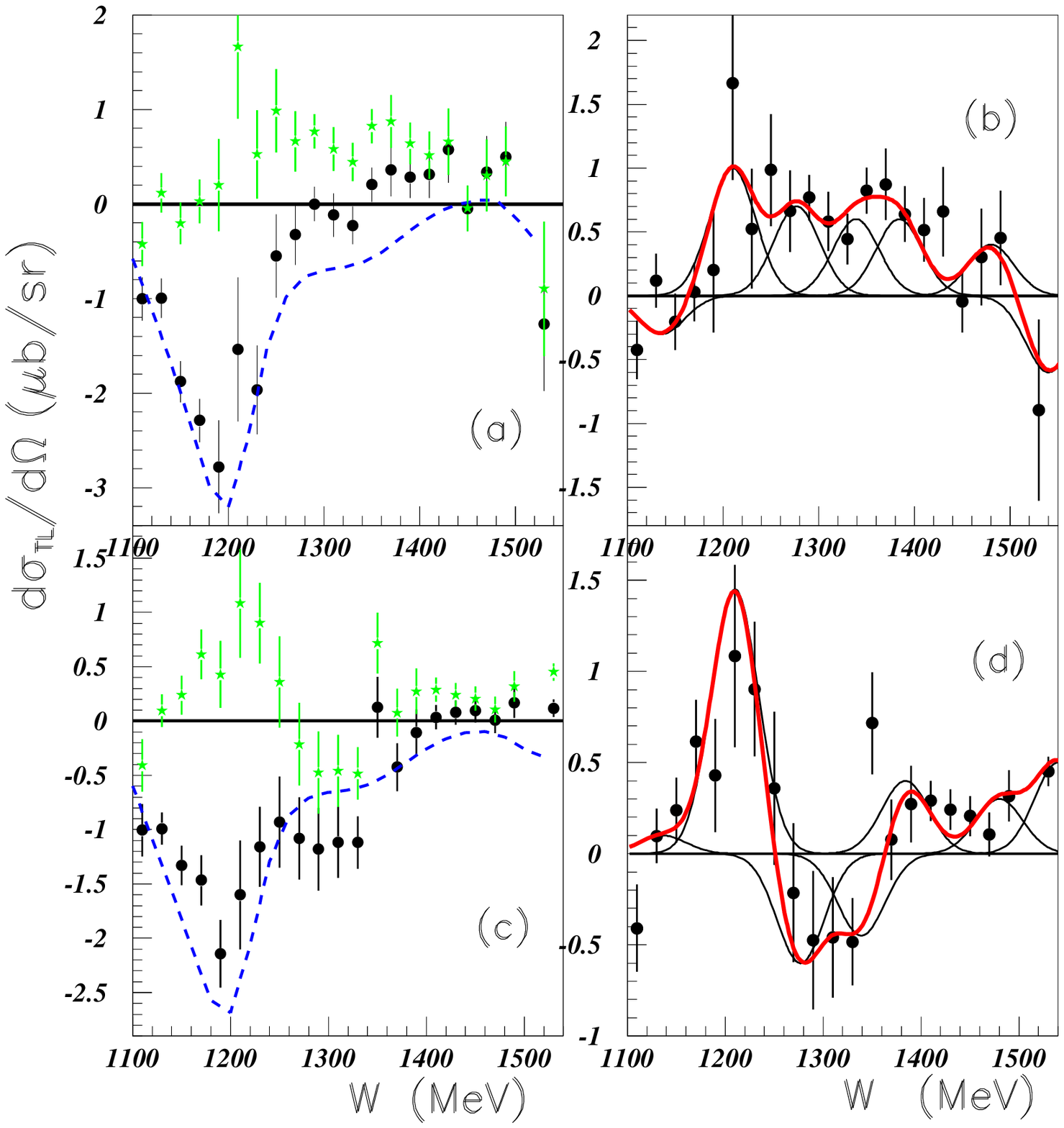}}
\caption [Fig.~31] {Same  as in Fig.~30, but for Q$^{2}$=0.4~GeV$^{2}$.}
\end{figure}
\end{center}

\begin{center}
\begin{figure}[!ht]
\scalebox{0.9}[0.9]{
\includegraphics[bb=12 12  530 550,clip,scale=1]{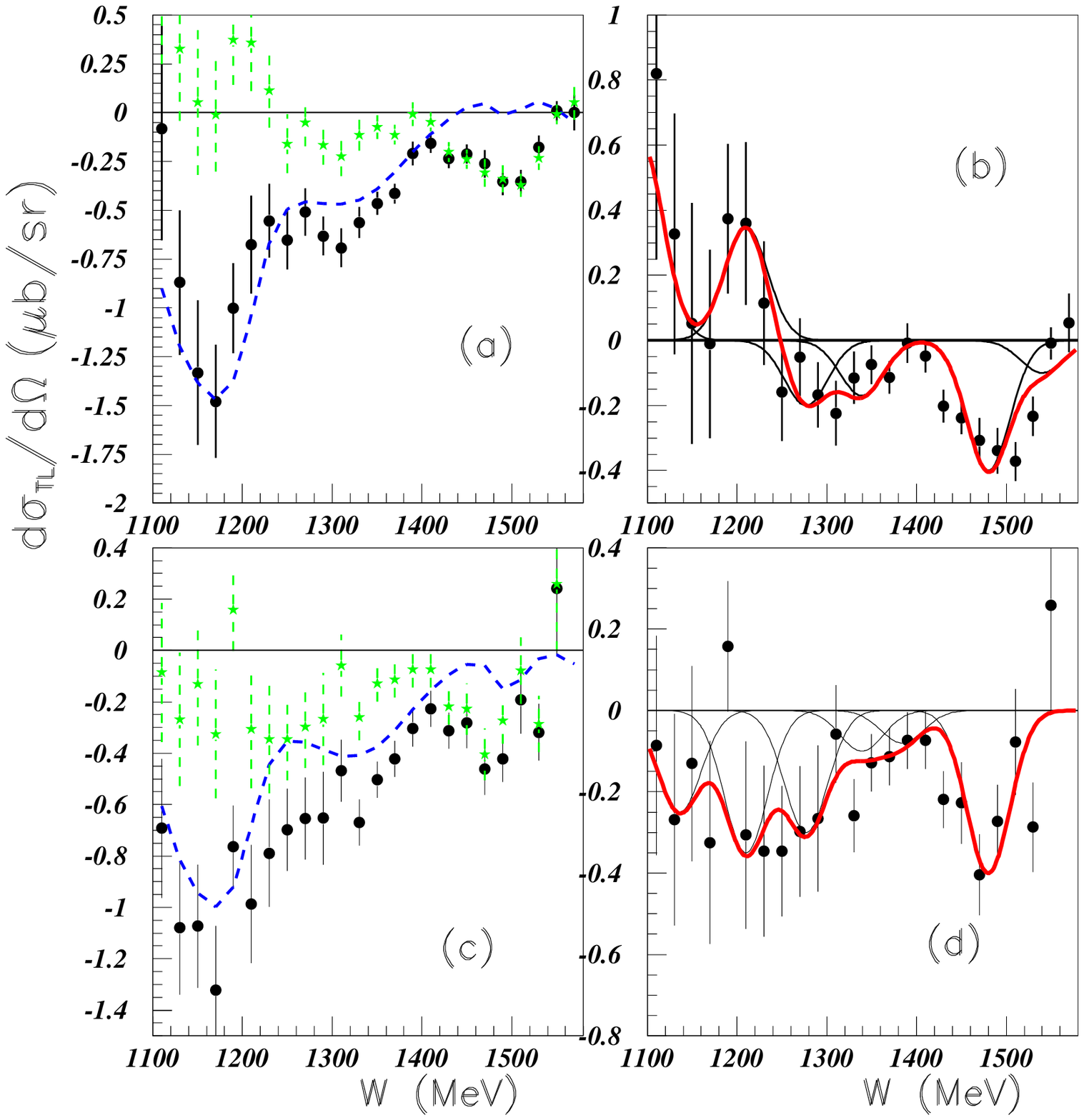}}
\caption [Fig.~32] {Same  as in Fig.~29, but for $\theta$=97.5$^{0}$.}
\end{figure}
\end{center}

\newpage\clearpage
\begin{center}
\begin{figure}[!ht]
\scalebox{0.9}[0.9]{
\includegraphics[bb=12 12  530 550,clip,scale=1]{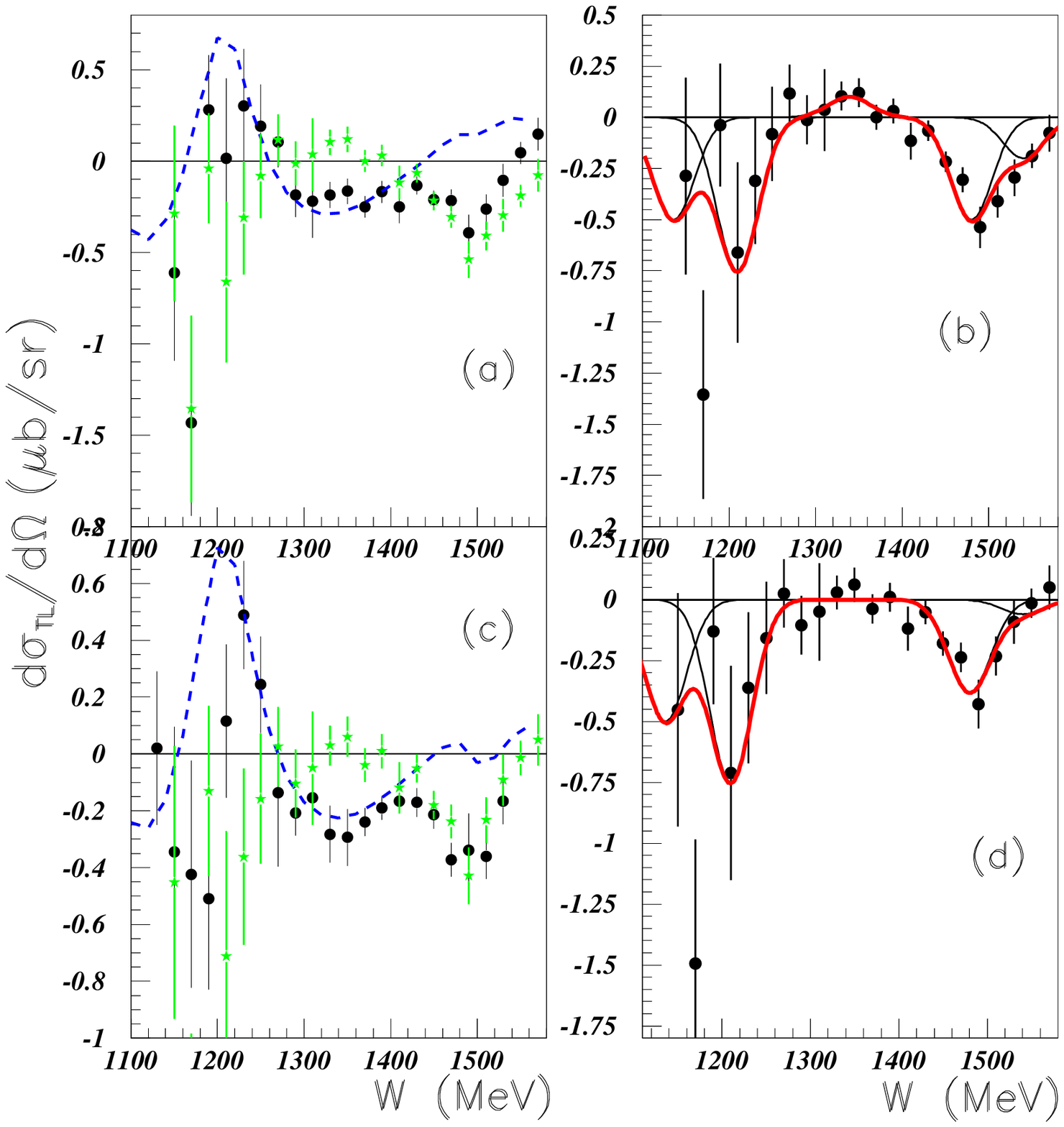}}
\caption [Fig.~33] {Same  as in Fig.~29, but for $\theta$=127.5$^{0}$.}
\end{figure}
\end{center}
\newpage\clearpage
\begin{center}
\begin{figure}[!ht]
\scalebox{0.9}[0.9]{
\includegraphics[bb=42 12  530 550,clip,scale=1]{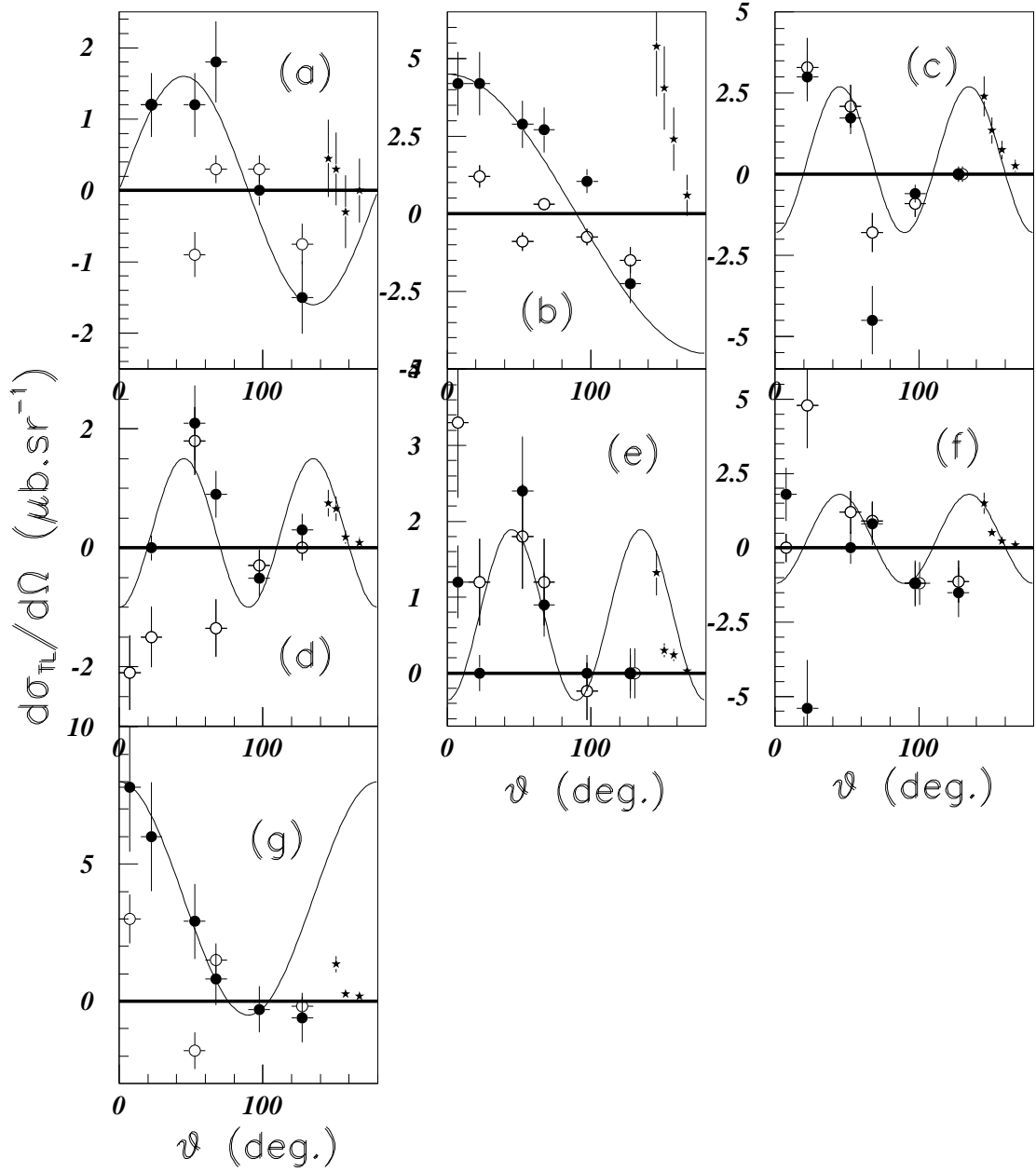}}
\caption [Fig.~34] {Same  as in Fig.~20 but showing the angular
variation of the cross-sections corresponding to the $\sigma_{TL}$ structure function for both reactions.}
\end{figure}
\end{center}

\small

\begin{table}
\begin{center}
\caption{Masses (in MeV) of narrow exotic baryons, observed previously in SPES3 data
and extracted from previous p($\alpha,\alpha$')X spectra measured at SPES4
\protect\cite{mor1} \protect\cite{mor2}.}
\label{Table 1}
\begin{tabular}{c c c c c c c c c c c c c} 
\hline
\small\hspace*{-14.mm}SPES3 mass&1004&1044&1094&1136&1173&1249&1277&1339&1384&&&1479\\
\small\hspace*{-14.mm}pic marker&(a)&(b)&(c)&(d)&(e)&(f)&(g)&(h)&(i)&(j)&(k)&(l)\\
\small\hspace*{-14.mm}SPES4 mass 0.8$^{0}$&&1052&1113&1142&1202&1234&1259& &1370&1394& &1478\\
\small\hspace*{-14.mm}SPES4 mass 2$^{0}$&996&1036&1104&1144&1198&1234&&1313&1370&&&1477\\
\hline
\hline
\small\hspace*{-14.mm}SPES3 mass&1505&1517&1533&1542&(1554)&1564&1577&&&&\\
\small\hspace*{-14.mm}pic marker&(m)&(n)&(o)&(p)&(q)&(r)&(s)&&&\\
\small\hspace*{-14.mm}SPES4 mass 2$^{0}$&1507&1517&1530&1544&1557&1569&1580&&&&\\
\hline
\end{tabular}
\end{center}
\end{table}

\begin{table}
\begin{center}
\caption{Tentative attribution of isospin, using figs.~20, 28, and 34, for the narrow structures (see text).  Inserts (a), (b), (c), (d), (e), (f), and (g) correspond
respectively to the following masses: M=1136~MeV, 1210~MeV,  1277~MeV, 1339~MeV, 1384~MeV, 1480~MeV, and 1540~MeV.}
\label{Table 2}
\begin{tabular}{c c c c c c c c} 
\hline
insert&(a)&(b)&(c)&(d)&(e)&(f)&(g)\\
\hline
$\sigma_{T} + \epsilon \sigma_{L}$&1/2&3/2&3/2&1/2&1/2&1/2&3/2\\
$\sigma_{TT}$&1/2&3/2&1/2&1/2&1/2&3/2&1/2\\
$\sigma_{TL}$&1/2&1/2&3/2&3/2&3/2&3/2&1/2\\
\hline
\end{tabular}
\end{center}
\end{table}

\end{document}